\documentclass[%
superscriptaddress,
preprint,
 amsmath,amssymb,
 aps,
prb,
]{revtex4-2}

\usepackage{graphicx}
\usepackage{dcolumn}
\usepackage{bm}
\usepackage{multirow}
\usepackage{amsmath}
\usepackage{ulem}
\usepackage{xcolor}

\usepackage{stackrel}
\usepackage{amssymb,amsfonts}

\begin{document}

\preprint{ver1}

\title{\textbf{Imaging van Hove Singularity Heterogeneity in Overdoped Graphene} 
}%

\author{Raymond Blackwell}
    \email{raymond.blackwell@stonybrook.edu}
 \affiliation{Department of Physics and Astronomy, Stony Brook University, Stony Brook, NY 11790 USA}%
 
\author{Zengyi Du}
\affiliation{Condensed Matter Physics and Materials Science Department, Brookhaven National Laboratory, Upton, NY, 11973 USA\\Department of Physics and Astronomy, Stony Brook University, Stony Brook, NY 11790 USA}

\author{Takuya Okugawa} 
\affiliation{Institut f\"ur Theorie der Statistischen Physik, RWTH Aachen, 
52056 Aachen, Germany and JARA - Fundamentals of Future Information Technology.\\Max Planck Institute for the Structure and Dynamics of Matter, Center for Free Electron Laser Science, 22761 Hamburg, Germany.}

\author{Asish K. Kundu}
\affiliation{Condensed Matter Physics and Materials Science Department, Brookhaven National Laboratory, Upton, NY, 11973 USA\\National Synchrotron Light Source II, Brookhaven National Laboratory, Upton, NY, 11973 USA}

\author{Zebin Wu}
\affiliation{Condensed Matter Physics and Materials Science Department, Brookhaven National Laboratory, Upton, NY, 11973 USA}

\author{Ilya Drozdov}
\affiliation{Condensed Matter Physics and Materials Science Department, Brookhaven National Laboratory, Upton, NY, 11973 USA}

\author{Angel Rubio}
\affiliation{Max Planck Institute for the Structure and Dynamics of Matter, Center for Free Electron Laser Science, 22761 Hamburg, Germany\\Center for Computational Quantum Physics, Simons Foundation Flatiron Institute, New York, NY 10010 USA\\Nano-Bio Spectroscopy Group, Departamento de Fisica de Materiales, Universidad del Pa\'is Vasco, UPV/EHU- 20018 San Sebasti\'an, Spain.}

\author{Dante M. Kennes}
\email{dante.kennes@rwth-aachen.de}
\affiliation{Institut f\"ur Theorie der Statistischen Physik, RWTH Aachen, 
52056 Aachen, Germany and JARA - Fundamentals of Future Information Technology\\Max Planck Institute for the Structure and Dynamics of Matter, Center for Free Electron Laser Science, 22761 Hamburg, Germany.}

\author{Kazuhiro Fujita} 
\email{kfujita@bnl.gov}
\affiliation{Condensed Matter Physics and Materials Science Department, Brookhaven National Laboratory, Upton, NY, 11973 USA}%

\author{Abhay N. Pasupathy} 
\email{apn2108@columbia.edu}
\affiliation{Condensed Matter Physics and Materials Science Department, Brookhaven National Laboratory, Upton, NY, 11973 USA\\Department of Physics, Columbia University, New York, NY 10027 USA}%

\date{\today}

\begin{abstract}
Tuning the chemical potential of a solid to the vicinity of a van Hove singularity (vHS) is a well-established route to discovering emergent quantum phases. 
In monolayer graphene, the use of electron-donating metal layers has recently emerged as a method to dope the chemical potential to the nearest vHS, as evidenced by Angle-Resolved Photoemission Spectroscopy (ARPES) measurements. 
In this work, we study the spatial uniformity of the doping from this process using spectroscopic imaging scanning tunneling microscopy (SI-STM). 
Using molecular beam epitaxy (MBE), we achieve electron doping of graphene on SiC using Ytterbium (Yb-Graphene). 
We show using in-situ ARPES that the chemical potential is shifted to within 250 meV of the vHS. 
Using in-situ SI-STM, we establish that there exists significant inhomogeneity in the vHS position in overdoped graphene. 
We find two separate reasons for this. 
First, the spatial inhomogeneity of the intercalated Yb leads to local variations in the doping, with a length scale of inhomogeneity set by the screening length of $\sim$ 3 nm. 
Second, we observe the presence of substitutional Yb dopants in the graphene basal plane. 
These Yb dopants cause a strong local shift of the doping, along with a renormalization of the quasiparticle amplitude. 
Theoretical calculations confirm that the Yb impurities effectively change the local potential, thus energetically shifting the position of the van Hove singularity. 
Our results point to the importance of considering the spatial structure of doping and its inextricable link to electronic structure.
\end{abstract}

\maketitle
\newpage

The susceptibility to form new electronic phases in solids is often enhanced by a large density of electronic states (DOS) at the Fermi level. 
Within two-dimensional (2D) materials, various methods including chemical doping \cite{liu_chemical_2011}, charge transfer \cite{CHEN2009279,BrusandKim, Brusa}, electrostatic doping \cite{KellyAndZhong,PhysRevLett.114.066803,WangAndZettl}, electric fields \cite{zhang_direct_2009} and twist angle \cite{bistritzer_moire_2011, PhysRevLett.121.026402} have been used to engineer the DOS at the Fermi level and realize emergent electronic phases. 
Methods such as charge transfer and electrostatic doping aim to simply shift the chemical potential towards a peak in the DOS - e.g. by a van Hove singularity (vHS). 
On the other hand, methods such as twisting and electric field tuning aim to change the electronic structure itself.
A long-standing material of interest in this regard is monolayer graphene, the paradigmatic 2D material. 
Pristine monolayer graphene has a Dirac point at the chemical potential, and two vHSs located approximately 2 eV away from the Dirac point (Figure 1A). 
Several theoretical predictions \cite{PhysRevLett.100.146404,nandkishore_chiral_2012,PhysRevLett.97.146401} have been made for interesting correlated electronic phases, including correlated insulating and superconducting states, if the chemical potential is brought close to these vHS. 

Shifting the chemical potential of monolayer graphene to one of the vHS requires doping the basal plane of graphene by $>10^{14}$ $cm^{-2}$, which is not possible by standard electrostatic doping due to gate dielectric failure. 
Several alternative methods have been attempted to achieve this, including charge transfer from alkali metals deposited on the surface \cite{PhysRevLett.108.156803}, ionic liquid gating \cite{PhysRevB.84.161412} and chemical substitution \cite{PasupathyScience}. 
The largest doping achieved by these methods is in the mid-$10^{13}$ $cm^{-2}$ range. Recently, promising progress has been achieved by intercalating metals into graphene grown on silicon carbide (SiC) \cite{Kim_2016, yurtsever_effects_2016, briggs_epitaxial_2019}. 
The metal atoms can diffuse between the graphene and SiC substrate and cause a large doping of the graphene by charge transfer \cite{RosenzweigPRB,RosenzweigPRL}. 
A myriad of atoms have been used for graphene intercalation but Lanthanide atoms seem to exert the highest levels of influence on the levels of electron doping \cite{pasti_atomic_2018,kim_manipulation_2017,SimonPhysRevB}.   

Previous studies have also shown that ytterbium intercalation is capable of doping epitaxial graphene beyond the $\pi^*$ vHS, leading to an extended vHS scenario reminiscent of high-$T_c$ cuprate superconductors \cite{IrkhinPhysRevLett,YudinPhysRevLett}. 
Furthermore, the intercalation of erbium has been shown to dope graphene up to the Lifshitz transition and induce a flat band \cite{ZaarourPhysRevResearch}.

\begin{figure}[htp]
\includegraphics[width=\columnwidth]{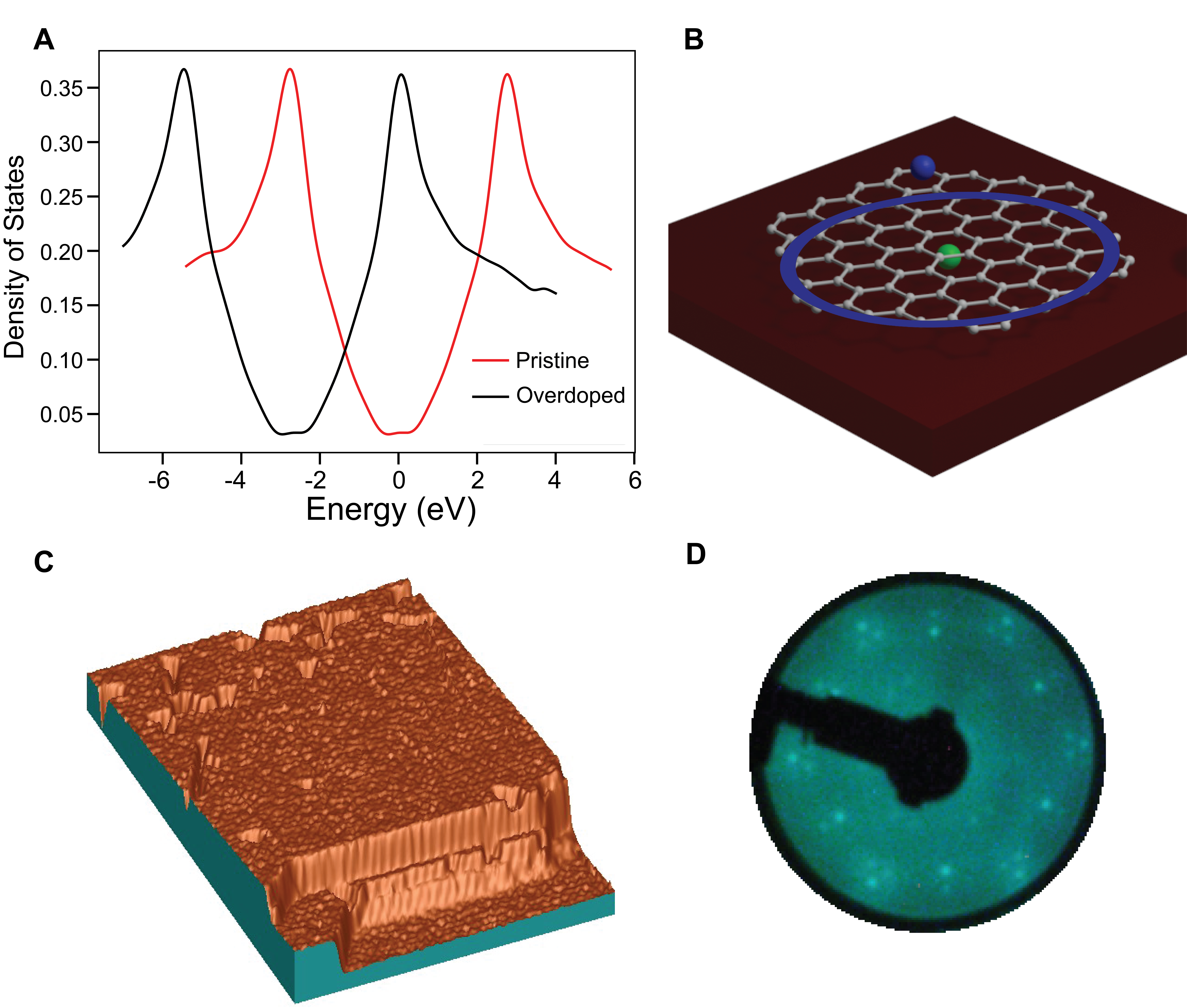}
\caption{Yb-Intercalated Quasi-Freestanding Monolayer Graphene (QFMLG)  (\textbf{A})~Calculated density of states for pristine (red) and overdoped graphene (black) showing the van Hove Singularity at the Fermi level.
(\textbf{B})~Cartoon schematic of two distinct types of Ytterbium dopants. The blue Ytterbium represents a substitutional dopant while the green Ytterbium is intercalated between the graphene and underlying silicon carbide. 
The blue circle surrounding the intercalated Ytterbium represents an approximate screening length.    
(\textbf{C})~Three-dimensional STM topograph of quasi-freestanding monolayer graphene (QFMLG) $V_{set}$ = 800~mV, $I_{set}$ = 400 pA.
(\textbf{D})~LEED diffraction pattern of QFMLG}
\label{fig1}
\end{figure}

While the achievement of these large doping levels in monolayer graphene is extremely interesting, much remains unknown about the electronic structure of these doped materials including whether correlated electronic phases emerge. 
Some evidence that these systems show effects beyond simple uniform doping exists already from ARPES, where the electronic band structure for the heavily doped system showed a reconstruction, possibly due to the coupling of electrons to localized states created by the intercalated atoms \cite{RosenzweigPRB,RosenzweigPRL}. 
One important unexplored aspect due to the limitations of ARPES measurements is the relationship between intercalated atoms and the possible structural and electronic inhomogeneity in graphene that can arise. 
Such effects are well known in lightly doped monolayer graphene, which exhibits an inhomogeneous distribution of charge that is caused by randomly distributed dopants, yielding charge puddles \cite{ZhangNaturePhys,martin_observation_2008} and scattering interference in their vicinity \cite{Stroscio}.  
Additionally, it is unknown if any of the deposited metal atoms can be incorporated into the graphene basal plane (Figure \ref{fig1}B). 
Theoretically, it is predicted that the electronic properties of pristine graphene can be dramatically modified by the incorporation of substitutional dopant atoms. 
For example, magnetic transitional metal impurities are believed to induce distinct magnetic moments in the graphene lattice \cite{crook_proximity-induced_2015}. 
Much less is known about the effects of substitutional Lanthanide atoms as a result of their complex electronic structure \cite{basiuk_adsorption_2022}. 
Experimental and theoretical works have shown evidence for enhanced electron-phonon and electron-electron interactions between hybridized ytterbium atoms and graphene \cite{HwangPRB,KANG20191}. 
It has additionally been posited that the adsorption of Lanthanide atoms can affect the C-C bond length in graphene and the spin state of the adsorbed atom. 
Therefore, it is especially important to characterize the effect of ytterbium intercalation on the local electronic structure.  

In this work, we use atomically-resolved STM measurements to address these questions. 
We complement these measurements with Low-energy Electron Diffraction (LEED) and ARPES measurements which probe the structure and spatially averaged electronic properties. 
Our samples are produced in an MBE chamber that is connected in-situ with the ARPES and STM chambers. 
This approach allows us to transfer the samples without breaking vacuum and enables us to study an identical surface by the different spectroscopic probes \cite{CKKim2018}. 
Following the procedure of Emstev \textit{et al.} \cite{GOLER2013249}, the carbon buffer layer was grown on 4H-SiC(0001), and subsequent ytterbium intercalation at a sample temperature of 800 K ultimately yields quasi-free-standing monolayer graphene (QFMLG). 
The surface was initially analyzed using LEED which reveals no evidence for ordered intercalation (Figure \ref{fig1}D). 
The LEED diffraction pattern shows only the SiC reconstruction relative to the topmost silicon atoms in the SiC substrate. 

\begin{figure}[htp]
\centering
\includegraphics[width = 5.25 in]{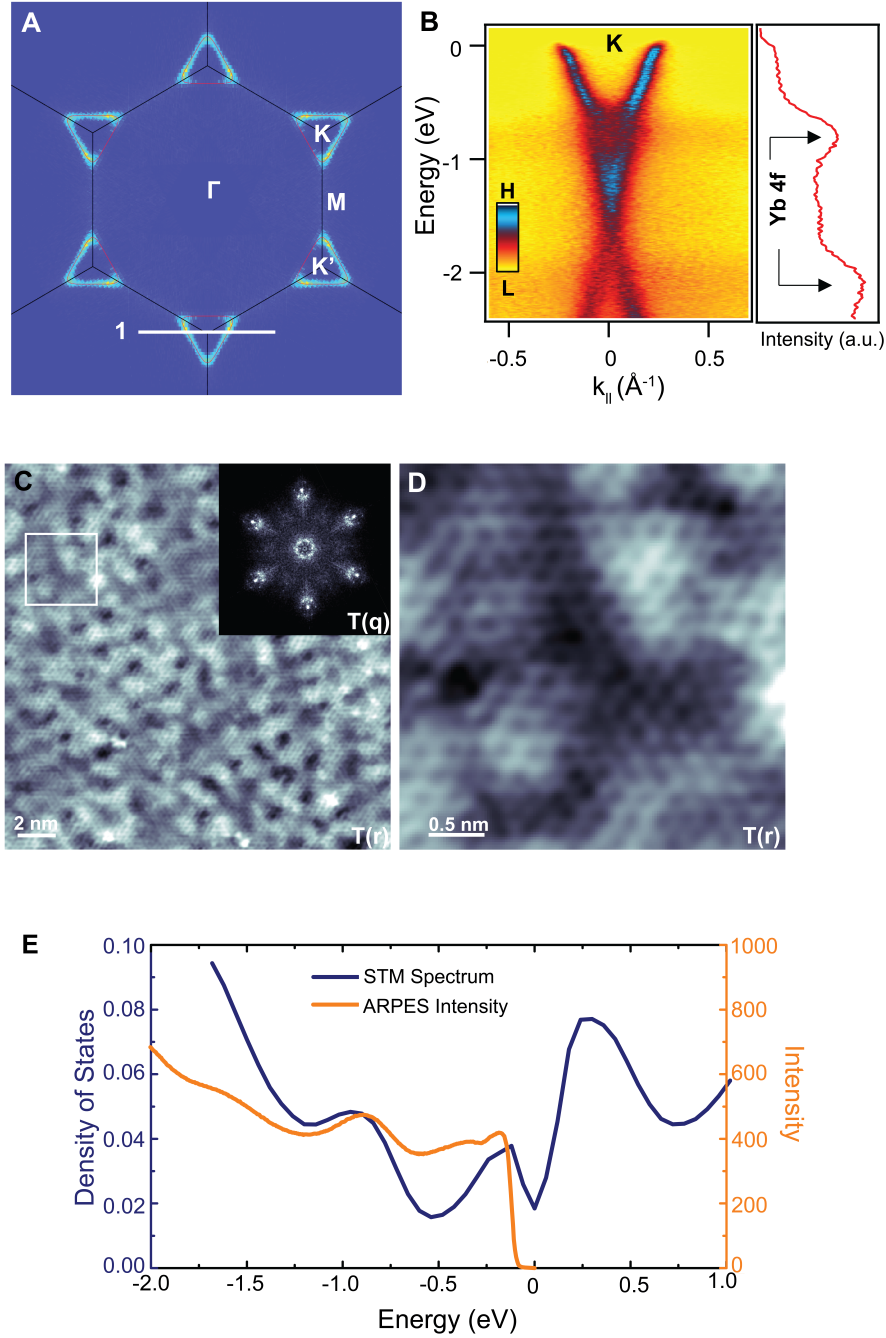}
\caption{Characterization of QFMLG on SiC  (\textbf{A})~STM topography of a uniform 20 nm FOV region Inset: Fourier transform showing Bragg peaks associated with graphene. 
$V_{set}$ = 400~mV, $I_{set}$ = 300 pA
(\textbf{B})~Zoom-in topograph of region in (a)  $V_{set}$ = -300~mV, $I_{set}$ = 200 pA. 
(\textbf{C})~Fermi surface of QFMLG determined by ARPES (\textbf{D})~High-resolution energy-momentum cuts along the path labeled 1 in (C) along with integrated ARPES intensity. 
(\textbf{E})~Comparison of integrated ARPES intensity to STM spectra averaged over the entire 20 nm FOV ($V_{set}$ = -400~mV, $I_{set}$ = 100 pA)}
\label{fig2}
\end{figure}

We study the surface using ARPES immediately after synthesis. 
The Fermi surface displayed in \ref{fig2}A shows that Ytterbium intercalation in the ($6\sqrt{3} \times 6\sqrt{3}$) carbon buffer layer leads to a carrier density of $\sim 2.4*10^{14} cm^{-2}$. 
Unlike the Fermi surface of pristine graphene with two triangular electron pockets centered on the $K$ and $K'$ points, there is substantial trigonal warping leading to increased intensity along the KMK' direction, which is an indication of an extended vHS. 
A high-resolution energy-momentum cut recorded at $K$ perpendicular to the $\Gamma$K direction along the white line labeled 1 is presented in Figure \ref{fig2}B. 
The band structure in Figure \ref{fig2}B is characterized by several notable features--- first, the kink roughly 200 meV below the $E_F$, typically attributed to the electron-phonon coupling \cite{zhang_giant_2008}. 
The next prominent feature is a pair of non-dispersive bands appearing $\sim$ 800 meV and 2 eV below the $E_F$, respectively, that are assigned to localized Yb 4$f$ states \cite{RosenzweigPRB}. 
The distortion of the conical graphene bands near the Yb states offer evidence of hybridization between the $\pi$ bands and the 4$f$ states, consistent with previous reports \cite{RosenzweigPRB}. Moreover, no replica bands or superstructures have been observed, indicating that the Yb atoms do not form an ordered structure. 
Finally, the Dirac point is identified roughly 1.4 eV below the $E_F$ confirming that the samples are heavily $n$ doped.

The QFMLG films was transferred into the STM $in-situ$ under ultra-high vacuum after the completion of ARPES measurements. 
All SI-STM measurements herein were conducted at $T = 8$ K using a tungsten tip that was first calibrated on an Au(111) surface. 
A three dimensional topographic image $T(r)$ of the as-grown QFMLG film can be seen in Figure \ref{fig1}C showing a surface with several holes and multiple terraces. 
As evidenced by the large area topographic image, the surface is highly inhomogenous, which is not ideal for SI-STM experiments. 
In order to obtain high quality data, it is necessary to focus on smaller areas such as the 23 nm $\times$ 23 nm field of view (FOV) in Figure \ref{fig2}C. 
Within this FOV, missing atoms (dark spots) and additional impurities (bright spots) are readily observed. 
The inset in Figure \ref{fig2}C is the Fourier transform $T(q)$ of the topographic image in Figure \ref{fig2}A and shows the Bragg peaks of graphene without additional order from the Yb dopant atoms, consistent with the LEED pattern in Figure \ref{fig1}D. 
Figure \ref{fig2}D represents a zoomed-in image of the area in Figure \ref{fig2}C marked by the white box. 
In this 5 nm $\times$ 5 nm FOV, the graphene lattice is clearly resolved, but there remains no evidence suggesting the positions of Yb atoms. 
In order to determine the location of Yb atoms, it is necessary to measure the local spectroscopic properties of QFMLG. 

\begin{figure}[htp]
    \centering
    \includegraphics[width=\columnwidth]{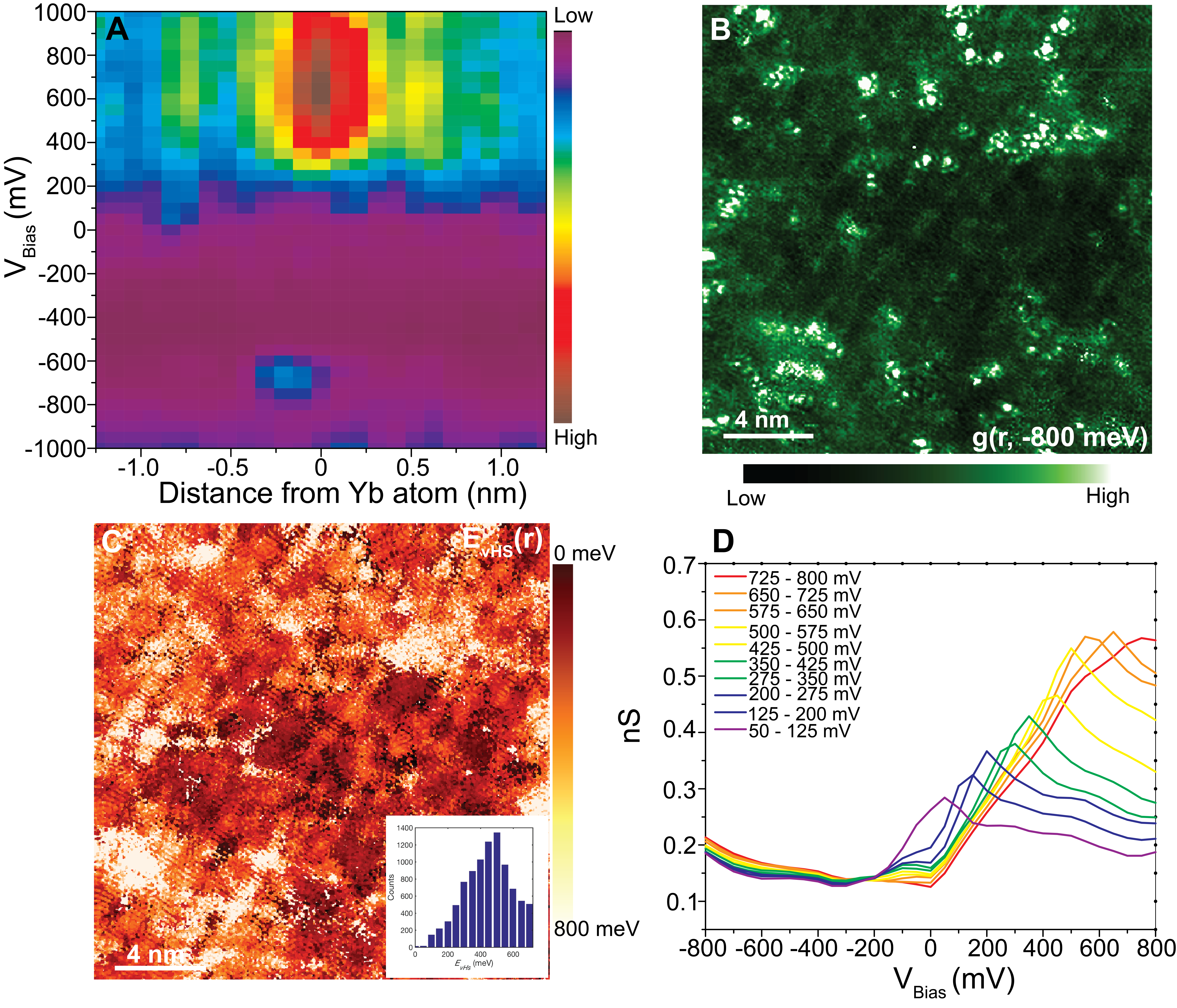}
    \caption{Inhonomgenous charge distribution.  (\textbf{A})~Experimentally measured dI/dV intensity map in the vicinity of substitutional Yb atom  (\textbf{B})~Differential conductance map, $g(r)$, taken at -800 meV, the energy associated with localized Yb states as determined by ARPES  
    (\textbf{C})~van Hove Singularity (vHS) map showing the peak position of the vHS over the entire 20 nm FOV. 
    Inset: Histogram of the distribution of peak positions (\textbf{D})~Gap sorted average spectra wherein individual spectrum are binned by the position of the vHS peak and averaged together. 
    STM setup conditions: $V_{set}$ = 200 mV, $I_{set}$ = 50 pA}
\label{fig3}
\end{figure}

In STM, the local density of states (LDOS) is visualized in the differential conductance spectrum, $g(r,E)\equiv dI/dV(r, E)$, thus permitting the acquisition of spatial LDOS maps as a function of bias voltage while simultaneously measuring the topography. 
Figure \ref{fig2}E contains the LDOS spectrum spatially averaged over the FOV in Figure \ref{fig2}A and compares this data to that obtained via ARPES. 
The data are in excellent agreement with each other and the peak in the LDOS spectrum at 400 meV is assigned to the vHS. 
All other peaks in the LDOS spectrum are reproduced from the ARPES energy distribution curves (Yb states: $-$0.8 eV, Dirac point: $-$1.4 eV). 
The positions of the Yb dopants are revealed by differential conductance mapping at the $-$800 meV, $g(r, E = -800 \textrm{meV})$, in Figure \ref{fig3}B. 
The bright white spots are localized impurity states that can be assigned to substitional Yb atoms as the number of Yb atoms in the FOV is not equal to the level of doping. 
A simple visual analysis of the bright white spots in Figure \ref{fig3}B shows there are $\sim$ 50 such spots in the 23 $\times$ 23 nm FOV. 
Given that the doping density calculated via ARPES is $\sim 10^{14} cm^{-2}$, these spots cannot constitute all of the Yb atoms. 
It is also the fact that a STM measurement on the un-doped graphene on SiC has shown no localized impurity states, suggesting that the silicon atoms in the substrate layer are not intermixed with the carbon atoms within the graphene. 
Taken in concert, this data suggests that there are Yb atoms in the basal plane of graphene (substitutional Yb) and Yb atoms that are intercalated between the graphene and the SiC. 
It is also evident from Figure \ref{fig3}B that the substitutional Yb atoms are randomly distributed within the 23 nm $\times$ 23 nm FOV. 

The spatial heterogeneity of the Yb atoms begets an obvious question--- how does the electronic structure evolve over the same region? 
To answer this question, we begin by analyzing the LDOS in the vicinity of a Yb atom. 
Figure \ref{fig3}A shows that the electronic structure is drastically altered when the tip is directly over the Yb atoms. 
To explore the effect of the Yb atoms on the vHS, the peak position of the vHS is extracted at every pixel in the same FOV shown in Figure \ref{fig3}B. 
The vHS map, $E_{vHS}(r)$, is presented in Figure \ref{fig3}C and it is readily apparent that there is sizeable variation in the $E_{vHS}(r)$ as a function of position $r$. 
In Figure \ref{fig3}C, regions with more white color correspond to a higher $E_{vHS}(r)$ and there are noticeable similarities between Figures \ref{fig3}B and C. 
Regions where the substitutional Yb atoms lie in Figure \ref{fig3}B (bright white spots) are also characterized by higher $E_{vHS}(r)$ (white spots) in Figure \ref{fig3}C. 
Conversely, in regions without substitutional Yb atoms (dark green) in Figure \ref{fig3}B, the $E_{vHS}(r)$ appears more homogeneous with relatively lower $E_{vHS}$ values.
The inset of Figure \ref{fig3}C shows the distribution of $E_{vHS}(r)$ values and all spectra within each bin are averaged together to produce Figure \ref{fig3}D. 

\begin{figure}[htp]
\centering
\includegraphics[width = \columnwidth]{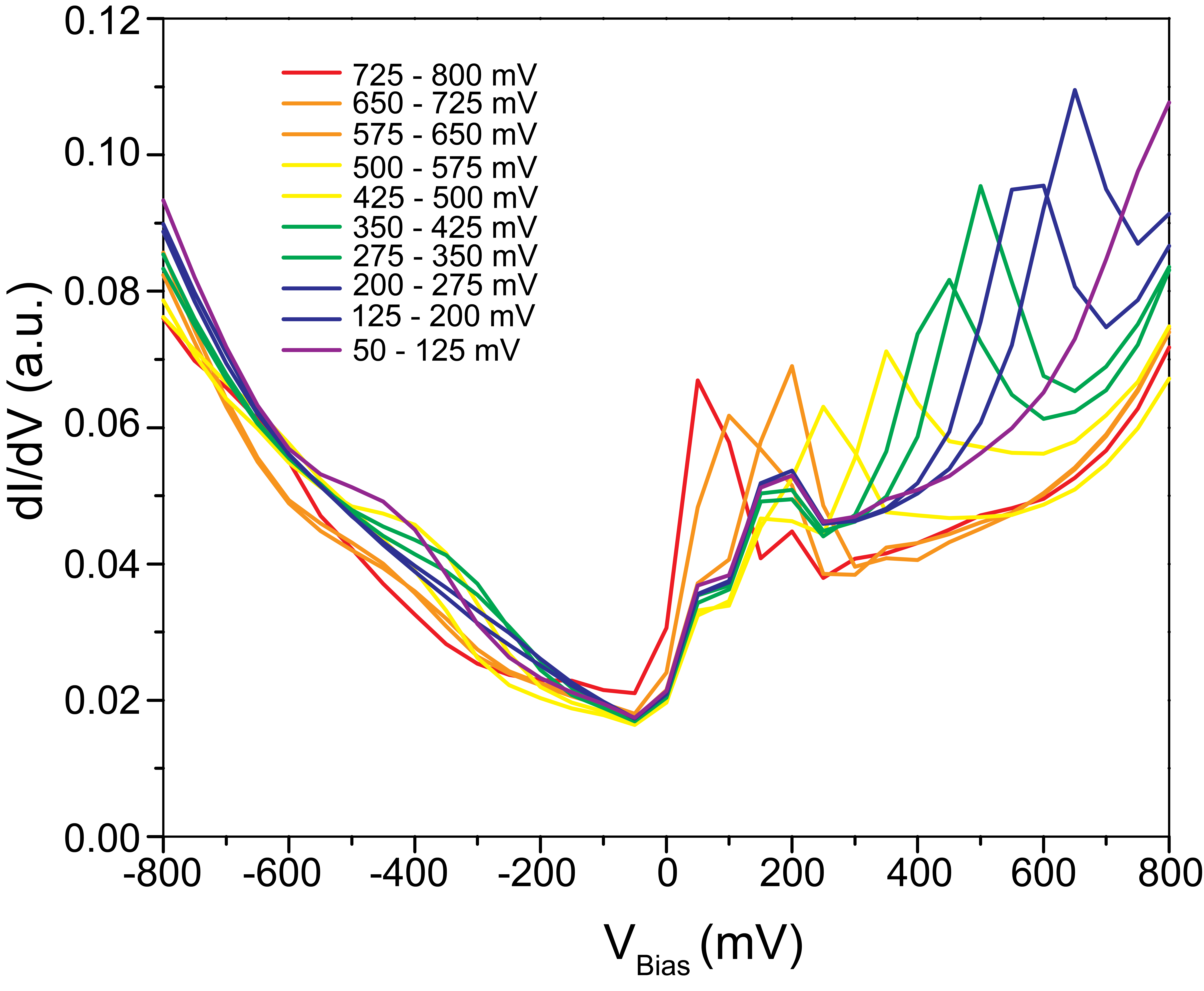}
\caption{{vHS-energy sorted dI/dV Spectra}  
~vHS-energy sorted average spectra wherein individual spectrum are binned by the position of the vHS peak and averaged together. 
STM setup conditions: $V_{set}$ = -200 mV, $I_{set}$ = 50 pA}
\label{Graph1}
\end{figure}

It is expected that in a rigid band shift picture, all of the spectra in Figure \ref{fig3}D would be shifted in energy as carriers are doped, and the peak amplitude would be roughly constant. 
While Figure \ref{fig3}C provides evidence that the vHS peak position varies spatially, Figure \ref{fig3}D shows that the peak amplitude is positively correlated with the peak position, inconsistent with the rigid band shift picture. 
Instead two distinct effects are observed--- first, the position of the vHS is altered as a direct consequence of the Yb inhomogenity. As described above, in regions without substitutional Yb atoms, the vHS moves towards the Fermi level consistent with previous ARPES studies where the vHS is driven to the Fermi level via Yb intercalation. 
Near the substitutional Yb atoms, this doping effect is locally screened with a characteristic length of $\sim$ 3 nm and the vHS is pushed away from the Fermi level. 
The reduction and broadening of the vHS peak is inconsistent with a rigid band shift picture, but instead points to interaction-related effects that come into play as more states become available near the Fermi level.  
Additional spectra were taken under positive bias setup conditions to rule out potential setup effects (Figure \ref{Graph1}. Here a similar trend was observed indicating this is an intrinsic feature of the Yb-intercalated graphene system. 

\begin{figure}[htp]
\centering
\includegraphics[width=\columnwidth]{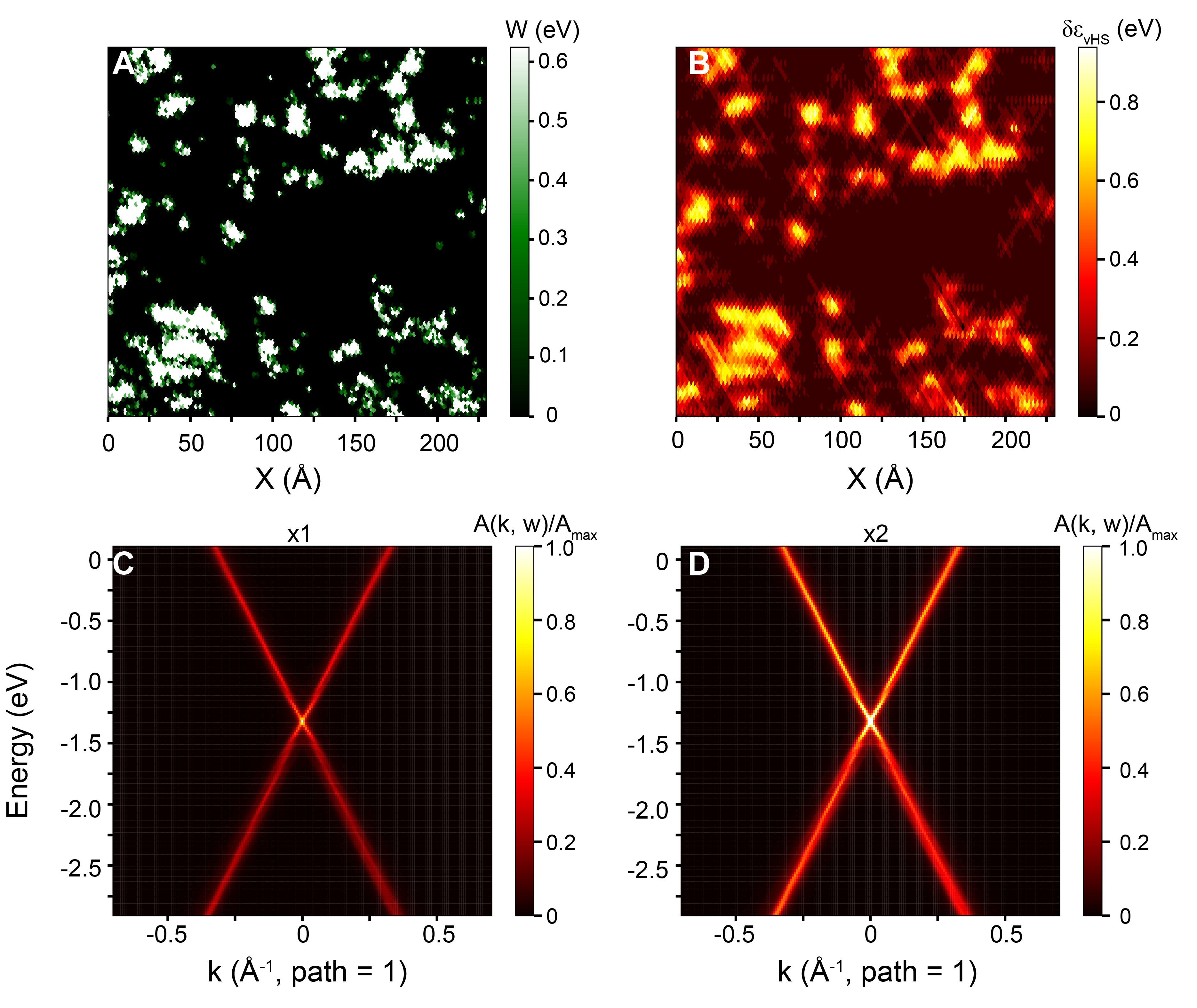}
\caption{Calculated Electronic Structure of QFMLG  
(\textbf{A})~The disorder profile generated by the local Gaussian potential, derived from the experimental conductance map shown in Figure \ref{fig3}B. (\textbf{B}) vHS shift produced by calculating the LDOS of the two dimensional graphene with the local Gaussian potential, where the spatial profile of the potential is given in panel \textbf{A}. (\textbf{C} and \textbf{D}) Momentum-resolved spectral functions for the two dimensional graphene with the local Gaussian potential shown in panel \textbf{A}. In panel \textbf{D}, the data intensity from panel \textbf{C} is multiplied by $2$ to highlight the non-zero broadening effect caused by the intercalated and substitutional Yb atoms. k(1) denotes the k-point path used for the ARPES measurement indicated in Figure \ref{fig2}A and B. The nearest neighbor hopping parameter $t$ is set to be $3$ eV. The parameters for the local Gaussian potential are $W=0.63$ eV and $\sigma=0.5a$ ($a$ is the carbon-carbon bond length). A finite strip consisting of $200 \times 200$ supercell is used for LDOS computation in panel \textbf{B}, while $82 \times 82$ supercell with periodic boundary condition is used for calculating the spectral function in panels \textbf{C} and \textbf{D}.}  

\label{fig4}
\end{figure}

In order to further corroborate the spatial variation in the vHS peak position, the experimental results are compared to theoretical calculations using the following method. 
First, the effect of substitutional and intercalated Yb atoms is modeled as a local Gaussian potential $W\exp{\left(-r^2/(2\sigma^2)\right)}$ where $W$ and $r$ represent the amplitude of the potential and the distance from the Yb atoms, respectively. $\sigma$ controls the spatial width of the potential. We use the same value of $W$($\sigma$) for both substitutional and intercalated Yb atoms. The position of the substitutional/intercalated Yb atom is identified from the differential conductance map presented in Figure \ref{fig3}B and the value of $W$ is estimated by assessing the maximum shift of LDOS (as in Figure \ref{fig1}A) created by the single Gaussian potential. The resultant disorder profile is shown in Figure \ref{fig4}A. 

We consider several values of $\sigma$ for the local Gaussian potential. Figure 5 shows the disorder profile and vHS shift $\delta \epsilon_{vHS}\equiv E_{vHS}(r)- \text{min}\left[E_{vHS}(r)\right]$ when $\sigma$ is half the spatial resolution of the differential conductance map in Figure \ref{fig3}B, where the potential is essentially a delta function. Figure 7 corresponds to $\sigma$ being equivalent to the graphene carbon-carbon bond length ($\sigma=a=1.42~\rm{\AA}$), and the case for $\sigma$ to be half of the bond length ($0.5a=0.71~\rm{\AA}$) is displayed in Figure \ref{fig4}.

We chose the intermediate value of $\sigma$ based on cross-correlation analysis between the simulated and experimental vHS maps presented in Figure 6 as the highest cross-correlation is obtained between theory and experiment when $\sigma$ is equal to $0.5a$. The magnitude of cross-correlation ($\sim$ 0.16) does not indicate a strong correlation as the subtleties in the experimental differential conductance map shown in Figure \ref{fig3}B are not captured in the disorder profile in Figure \ref{fig4}A.

\begin{figure*}[htp]
\centering
\includegraphics[scale = 0.5]{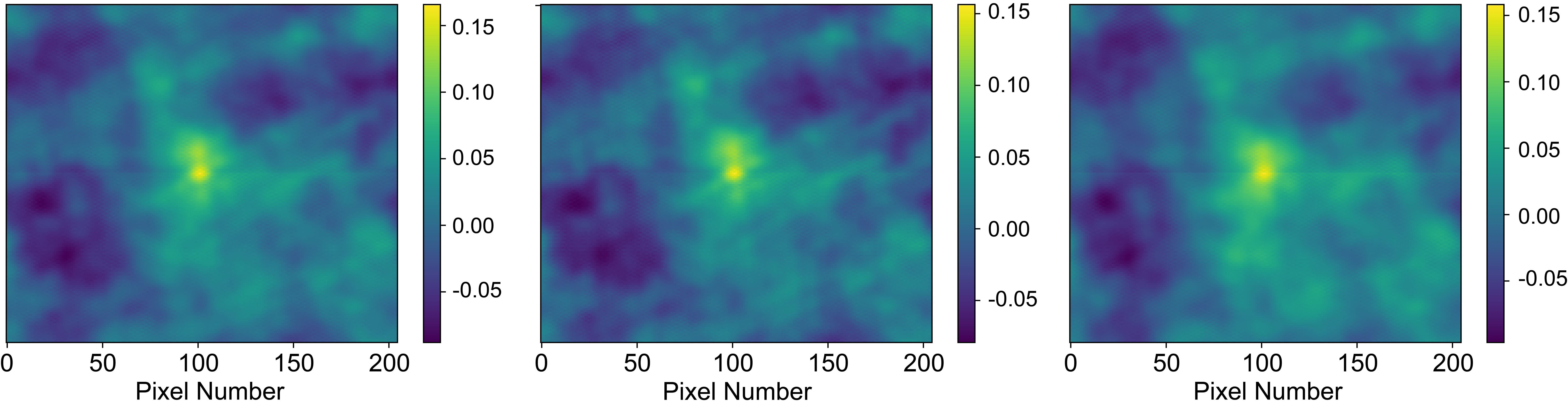}
\caption{Cross-correlation between experimental and calculated vHS when (A) $\sigma$ is equivalent to half the spatial resolution of the differential conductance map in Figure 3B, (B) $\sigma = 0.5a$, and  (C) $\sigma = a$.}
\label{Cross_Correlation}
\end{figure*}

The vHS shift $\delta \epsilon_{vHS}$ in Figure \ref{fig4}B  is produced by calculating the LDOS for the two dimensional tight-binding model~\cite{Reich2002} with the local Gaussian potential introduced above and identifying the vHS shift in the resultant LDOS (see Appendix for computational details). The overall feature of the experimentally observed vHS map shown in \ref{fig3}C is well captured by the computed vHS shift in Figure \ref{fig4}B.  

Having shown that the simple theoretical approach above reproduces key aspects of the STM data, we use the same approach introduced above to find the momentum-resolved spectral functions. To calculate the spectral function $A(\bm{k}, \omega)$ at a given momentum $\bm{k}$ and energy $\omega$ corresponding to the window of the ARPES measurement in Figure \ref{fig2}D, we apply the periodic boundary condition for all boundaries of the above tight-binding model and compute the spectral function for each momentum $\bm{k}$ and energy $\omega$ (see Appendix for computational details). The spectral function is unfolded by the same band structure unfolding procedure \cite{Wei2010, Popescu2012, Farjam2015}. The resultant spectral function is shown in Figure \ref{fig4}C while Figure \ref{fig4}D shows a magnified spectral function to highlight the broadening effects from the Yb atoms. The trend of broadening found in the experimental ARPES data in Figure \ref{fig2}D can be observed in Figure \ref{fig4}D emphasizing the utility of the model.    
These results suggest that there are two interconnected forms of inhomogenity induced via Yb atoms. 
One set of Yb atoms is intercalated between the SiC substrate and the quasi-freestanding monolayer graphene. 
These intercalated Yb atoms heavily n-dope the graphene such that the vHS is pushed towards the Fermi level. 
At the same time, we are able to locate Yb atoms that appear in the graphene basal plane, break translational symmetry, and also push the vHS away from the Fermi level.  
The occurrence of substitutional Yb atoms is unexpected given the stability of the graphene lattice. 
Typical strategies to produce subsitutional dopants in graphene involve post facto implantation via low-energy ions/plasma \cite{willke_doping_2015,BERTOTI2015185,RYBIN2016196} or incorporation during the graphene growth \cite{wei_synthesis_2009,PasupathyScience,schiros_connecting_2012}. 
In this work substitutional Yb atoms are simultaneously seen with intercalated Yb atoms and a similar effect has been observed with nitrogen atoms in epitaxial graphene on SiC through thermal reactions with ammonia \cite{wang_simultaneous_2013}.
The presence of two dissimilar Yb atoms directly leads to the observed local inhomogenity in the position of the vHS as can be seen in Figure \ref{fig3}C. 
It is readily apparent that there is a significant correlation between regions with substitutional Yb atoms (bright circles in Figure \ref{fig3}B) and regions where the vHS is further away from the Fermi level (bright patches in Figure \ref{fig3}C). 
In the areas without substitutional Yb atoms (dark regions in Figure \ref{fig3}B), the vHS map appears more homogeneous and the vHS is closer to the Fermi level. 
An interesting question to be explored in future work involves determining if any emergent quantum phases can arise within these nanoscale regions. 
Moreover, a newly developed technique to characterize local bond length variations in cuprates \cite{PhysRevX.13.021025} can be applied to this system to extract the gauge field in graphene in the vicinity of the substitutional Yb atoms. 

\begin{acknowledgments}
We thank Laura Classen and Robert Konik for discussions.
Work at Brookhaven is supported by the Office of Basic Energy Sciences, Materials Sciences and Engineering Division, U.S. Department of Energy under Contract No. DE-SC0012704. 
T. O., D.M.K, and A.R acknowledge support from the Max Planck-New York City Center for Non-Equilibrium Quantum Phenomena, the European Research Council, the Cluster of Excellence ‘Advanced Imaging of Matter' (AIM), Grupos Consolidados (IT1453-22), and the Deutsche Forschungsgemeinschaft (DFG, German Research Foundation) within the Priority Program SPP 2244 ``2DMP'' -- 443274199. 
The Flatiron Institute is a division of the Simons Foundation.
Computations were performed with computing resources granted by RWTH Aachen University under project rwth1280 as well as the HPC system Raven and Viper at the Max Planck Computing and Data Facility.
\end{acknowledgments}

\section*{Author's contribution}
Z.W. and I. D. synthesized the Yb-Graphene sample. A. K. performed ARPES measurements. Z. D., R. B. and K. F. performed SI-STM measurements. T. O. performed theoretical calculations. D.M.K. and A.R. supervised the theoretical calcuations. R. B., K. F., T. O. and A. N. P wrote the manuscript. R.B., Z.D., and T.O. contributed equally.

\section*{Competing Interests}
The authors declare no competing interests.

\section{Experimental Methods}
\subsection{Synthesis of the Yb-intercalated Graphene on SiC}
Yb-intercalated graphene was synthesized within the MBE module of the OASIS (OMBE-ARPES-SISTM) system in Brookhaven National Laboratory. Yb was evaporated from a Knudsen cell and deposited onto a 4H-SiC substrate terminated with a carbon buffer layer held at 250$^{\circ}$ C. The deposition rate of Yb was calibrated using a quartz crystal microbalance before the synthesis in the MBE chamber. Following deposition of Yb, the sample was annealed to 450$^{\circ}$ C to promote intercalation.
\subsection{ARPES measurements}
ARPES measurements were performed using a Scienta VUV5K microwave-driven plasma discharge lamp as a UV-photon source and a commercial Scienta R4000 electron spectrometer. The total experimental energy resolution was $\sim$ 20 meV during the measurements. Angular resolution was better than $\sim$ 0.15 deg and 0.5 deg along and perpendicular to the slit of the analyzer, respectively.
\subsection{SI-STM measurements}
SI-STM measurements were performed in a home-built STM with a base temperature of 8 K. Setup conditions for all experiments are indicated within the respective figure captions. 
\section{Theoretical Model Description}
For theoretical analysis, first, we map the experimental disorder profile shown in Figure 3B onto our theoretical lattice model shown in Figure 5A. Then, we model an effect of substitutional and intercalated Yb atoms as a local Gaussian potential $W\exp{\left(-r^2/(2\sigma^2)\right)}$ where $W$ and $r$ represent the amplitude of the potential and the distance from the Yb atoms, respectively. $\sigma$ controls the spatial width of the potential. Specifically, we use $W=0.63$~eV and three different $\sigma$ values; $\sigma$ being equivalent to half the spatial resolution of the differential conductance map in Figure 3B (essentially a delta function), half the graphene carbon-carbon bond length ($\sigma=0.5a$), and the graphene carbon-carbon bond length ($\sigma=a$).  

\begin{figure}[htp]
\centering
\includegraphics[scale = 0.4]{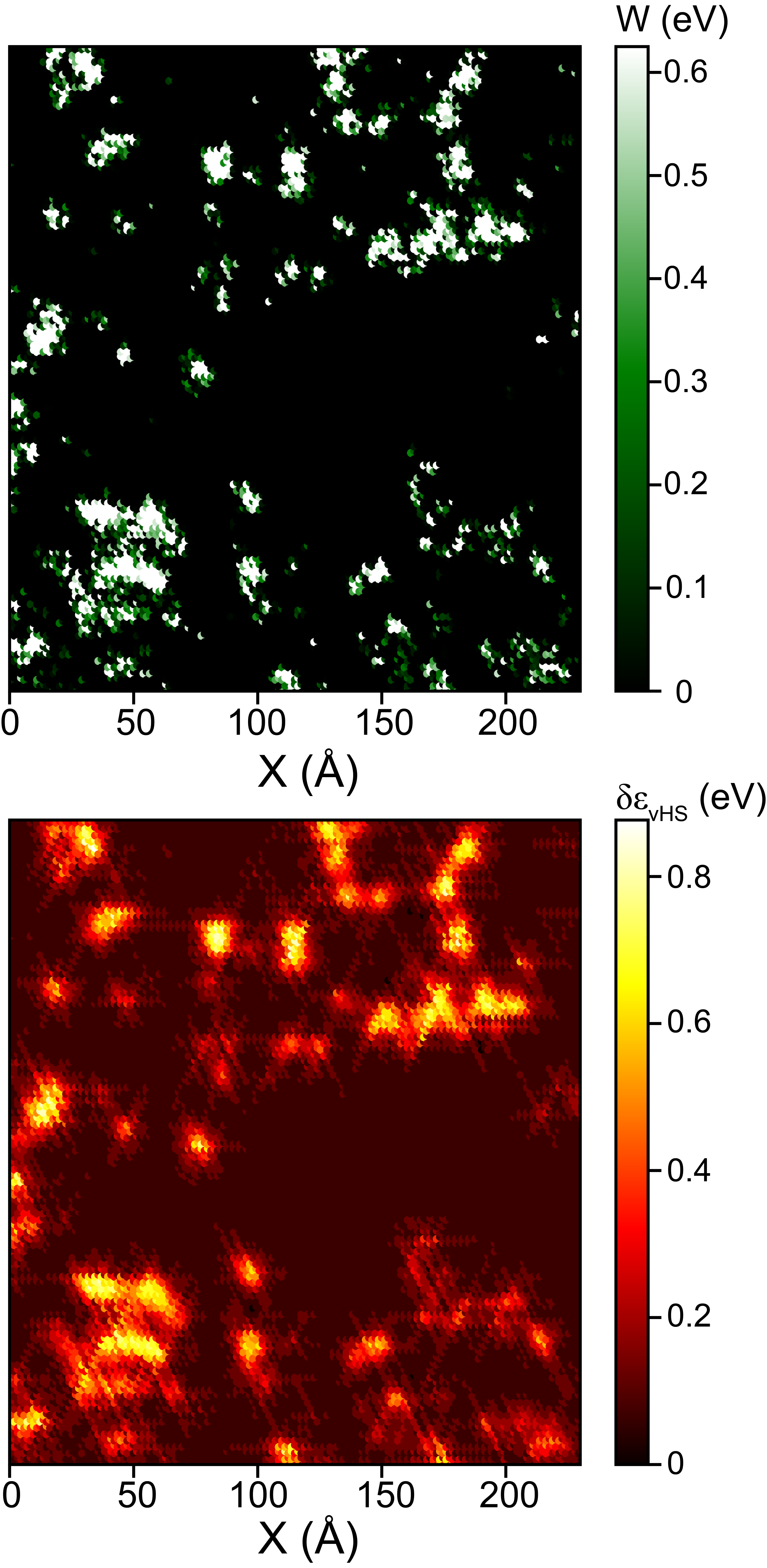}
\caption{The disorder profile and vHS shift as is introduced in Figure 5\textbf{A} and \textbf{B}. All parameters are the same as in Figure 5\textbf{A} and \textbf{B} except for $\sigma$ being equivalent to half the spatial resolution of the differential conductance map in Figure 3B}
\label{Sigma0}
\end{figure}

\begin{figure}[htp]
\centering
\includegraphics[scale = 0.4]{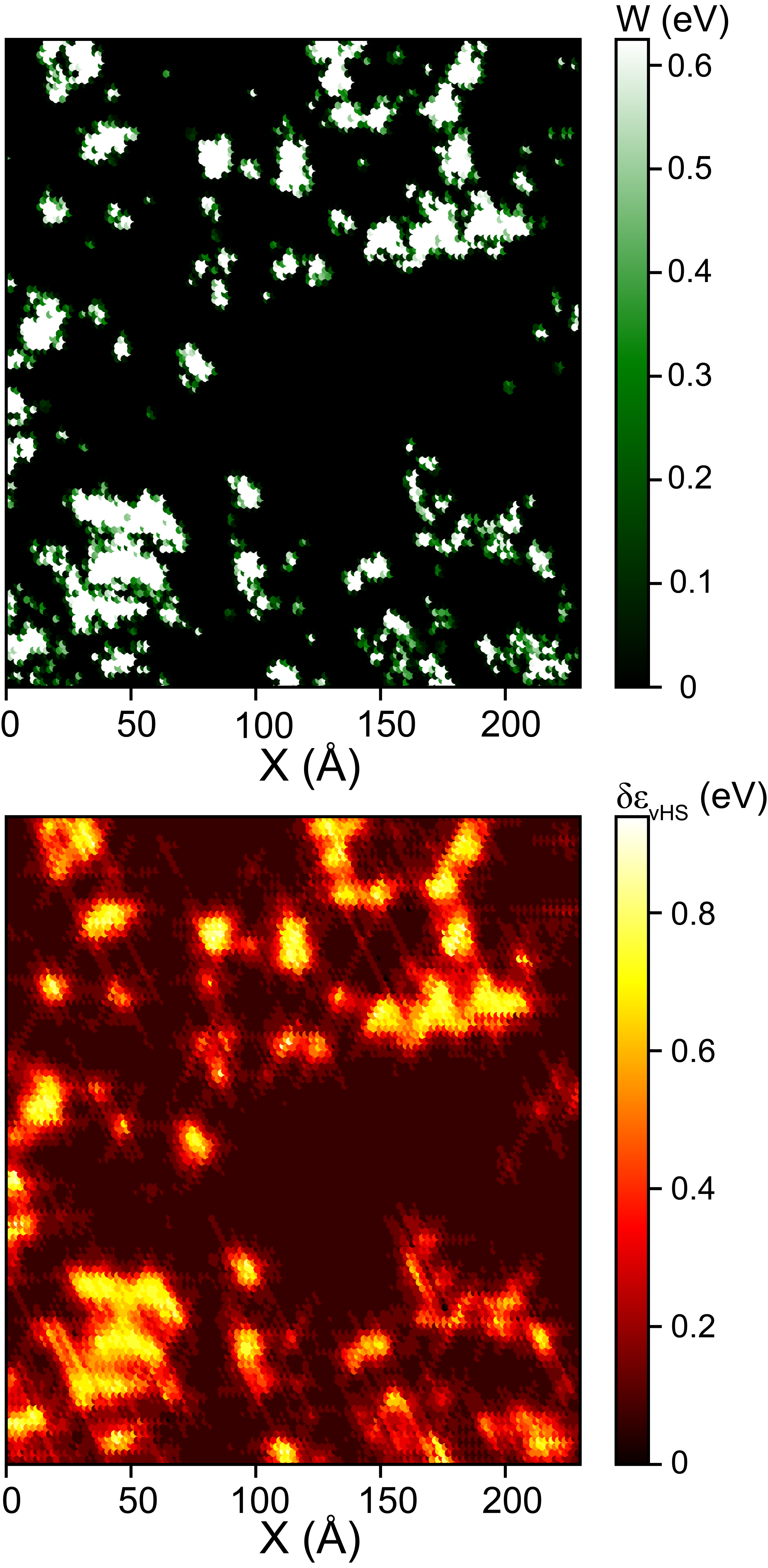}
\caption{The disorder profile and vHS shift as is introduced in Figure 5\textbf{A} and \textbf{B}. All parameters are the same as in Figure 5\textbf{A} and \textbf{B}.
}
\label{Sigma05}
\end{figure}

\begin{figure}[htp]
\centering
\includegraphics[scale = 0.4]{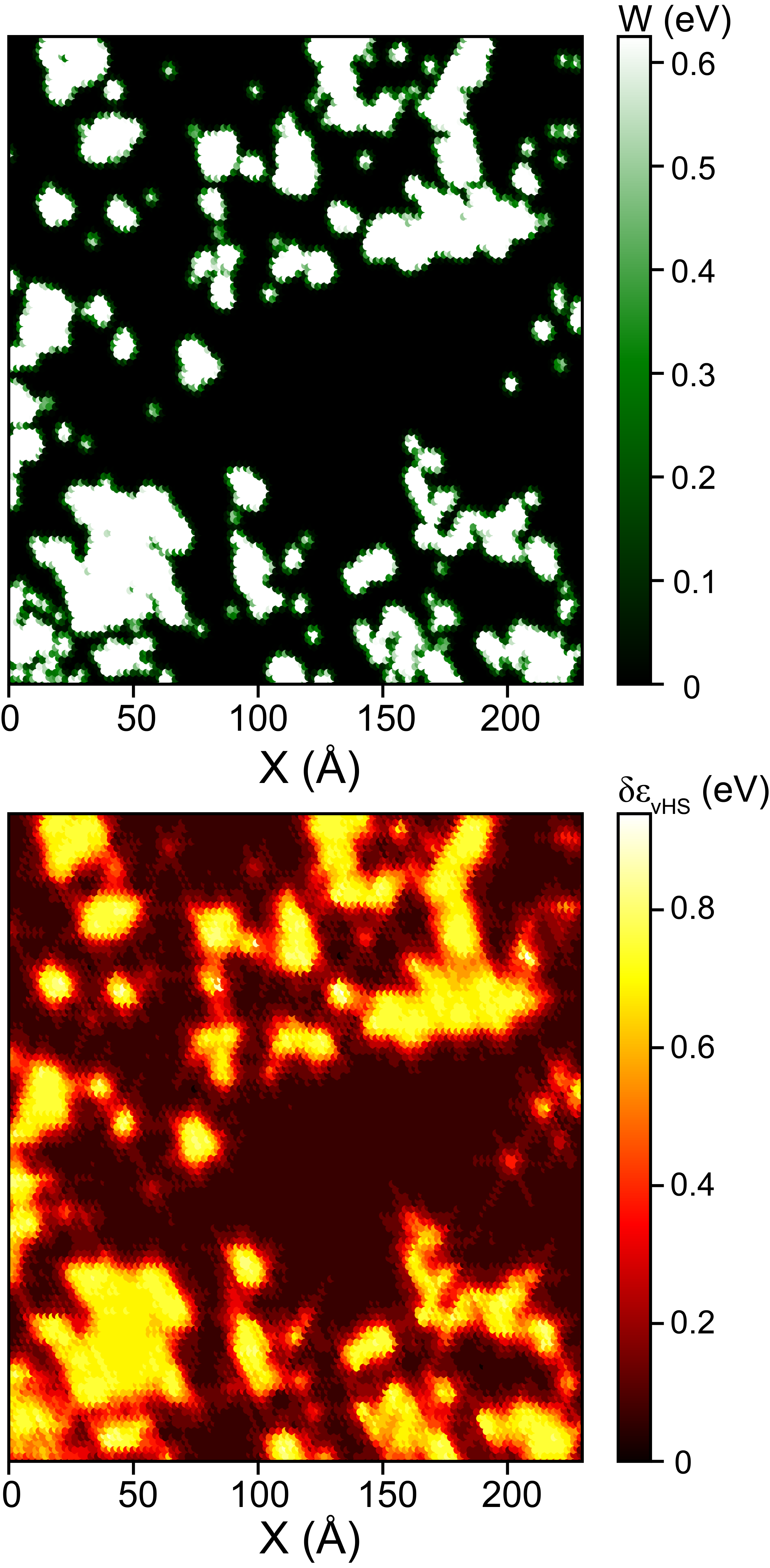}
\caption{The disorder profile and vHS shift as is introduced in Figure 5\textbf{A} and \textbf{B}. All parameters are the same as in Figure 5\textbf{A} and \textbf{B} except for $\sigma = a$.
}
\label{Sigma1}
\end{figure}

To produce the van Hove Singularity shift $\delta \epsilon_{vHS}$ shown in Figure 5B, we calculate the local density of states for the two dimensional graphene with substitutional/intercalated Yb and identify the van Hove Singularity in the resultant local density of states. We employ a real space large scale tight-binding model with a size of $200 \times 200$ unit cells to describe the two dimensional graphene with substitutional/intercalated Yb sample. To take an inverse of a large tight-binding matrix for computing the retarded Green's function, we employ a property of block tridiagonal matrix~\cite{Godfrin_1991}.

To obtain the momentum-resolved spectral function shown in Figure 5C and D, we employ the same tight-binding model and the local Gaussian potential to describe the two dimensional graphene with substitutional/intercalated Yb atom, yet with periodic boundaries and smaller system size due to computational expense. We take a supercell consisting of $82 \times 82$ unit cells which cover some of the substitutional/intercalated Yb profile shown in Figure 5A, upper panel of Figure 7, 8, and 9. On this finite strip, we set the periodic boundary condition for all boundaries. To calculate the spectral function $A(\bm{k}, \omega)$ on a given momentum $\bm{k}$ and energy $\omega$ corresponding to the window of ARPES measurement in Figure 2D, we unfold the spectral function by the same procedure as band structure unfolding~\cite{Wei2010, Popescu2012, Farjam2015}. The resultant spectral function is shown in Figure 5C and D.   

\bibliography{Graphene-Yb-vHS-ref}

\providecommand{\noopsort}[1]{}\providecommand{\singleletter}[1]{#1}
\begin{thebibliography}{49}%
\makeatletter
\providecommand \@ifxundefined [1]{%
 \@ifx{#1\undefined}
}%
\providecommand \@ifnum [1]{%
 \ifnum #1\expandafter \@firstoftwo
 \else \expandafter \@secondoftwo
 \fi
}%
\providecommand \@ifx [1]{%
 \ifx #1\expandafter \@firstoftwo
 \else \expandafter \@secondoftwo
 \fi
}%
\providecommand \natexlab [1]{#1}%
\providecommand \enquote  [1]{``#1''}%
\providecommand \bibnamefont  [1]{#1}%
\providecommand \bibfnamefont [1]{#1}%
\providecommand \citenamefont [1]{#1}%
\providecommand \href@noop [0]{\@secondoftwo}%
\providecommand \href [0]{\begingroup \@sanitize@url \@href}%
\providecommand \@href[1]{\@@startlink{#1}\@@href}%
\providecommand \@@href[1]{\endgroup#1\@@endlink}%
\providecommand \@sanitize@url [0]{\catcode `\\12\catcode `\$12\catcode
  `\&12\catcode `\#12\catcode `\^12\catcode `\_12\catcode `\%12\relax}%
\providecommand \@@startlink[1]{}%
\providecommand \@@endlink[0]{}%
\providecommand \url  [0]{\begingroup\@sanitize@url \@url }%
\providecommand \@url [1]{\endgroup\@href {#1}{\urlprefix }}%
\providecommand \urlprefix  [0]{URL }%
\providecommand \Eprint [0]{\href }%
\providecommand \doibase [0]{https://doi.org/}%
\providecommand \selectlanguage [0]{\@gobble}%
\providecommand \bibinfo  [0]{\@secondoftwo}%
\providecommand \bibfield  [0]{\@secondoftwo}%
\providecommand \translation [1]{[#1]}%
\providecommand \BibitemOpen [0]{}%
\providecommand \bibitemStop [0]{}%
\providecommand \bibitemNoStop [0]{.\EOS\space}%
\providecommand \EOS [0]{\spacefactor3000\relax}%
\providecommand \BibitemShut  [1]{\csname bibitem#1\endcsname}%
\let\auto@bib@innerbib\@empty
\bibitem [{\citenamefont {Liu}\ \emph {et~al.}(2011)\citenamefont {Liu},
  \citenamefont {Liu},\ and\ \citenamefont {Zhu}}]{liu_chemical_2011}%
  \BibitemOpen
  \bibfield  {author} {\bibinfo {author} {\bibfnamefont {H.}~\bibnamefont
  {Liu}}, \bibinfo {author} {\bibfnamefont {Y.}~\bibnamefont {Liu}},\ and\
  \bibinfo {author} {\bibfnamefont {D.}~\bibnamefont {Zhu}},\ }\bibfield
  {title} {\bibinfo {title} {Chemical doping of graphene},\ }\href
  {https://doi.org/10.1039/C0JM02922J} {\bibfield  {journal} {\bibinfo
  {journal} {J. Mater. Chem.}\ }\textbf {\bibinfo {volume} {21}},\ \bibinfo
  {pages} {3335} (\bibinfo {year} {2011})},\ \bibinfo {note} {publisher: The
  Royal Society of Chemistry}\BibitemShut {NoStop}%
\bibitem [{\citenamefont {Chen}\ \emph {et~al.}(2009)\citenamefont {Chen},
  \citenamefont {Qi}, \citenamefont {Gao},\ and\ \citenamefont
  {Wee}}]{CHEN2009279}%
  \BibitemOpen
  \bibfield  {author} {\bibinfo {author} {\bibfnamefont {W.}~\bibnamefont
  {Chen}}, \bibinfo {author} {\bibfnamefont {D.}~\bibnamefont {Qi}}, \bibinfo
  {author} {\bibfnamefont {X.}~\bibnamefont {Gao}},\ and\ \bibinfo {author}
  {\bibfnamefont {A.~T.~S.}\ \bibnamefont {Wee}},\ }\bibfield  {title}
  {\bibinfo {title} {Surface transfer doping of semiconductors},\ }\href
  {https://doi.org/https://doi.org/10.1016/j.progsurf.2009.06.002} {\bibfield
  {journal} {\bibinfo  {journal} {Progress in Surface Science}\ }\textbf
  {\bibinfo {volume} {84}},\ \bibinfo {pages} {279} (\bibinfo {year}
  {2009})}\BibitemShut {NoStop}%
\bibitem [{\citenamefont {Jung}\ \emph {et~al.}(2009)\citenamefont {Jung},
  \citenamefont {Kim}, \citenamefont {Jockusch}, \citenamefont {Turro},
  \citenamefont {Kim},\ and\ \citenamefont {Brus}}]{BrusandKim}%
  \BibitemOpen
  \bibfield  {author} {\bibinfo {author} {\bibfnamefont {N.}~\bibnamefont
  {Jung}}, \bibinfo {author} {\bibfnamefont {N.}~\bibnamefont {Kim}}, \bibinfo
  {author} {\bibfnamefont {S.}~\bibnamefont {Jockusch}}, \bibinfo {author}
  {\bibfnamefont {N.~J.}\ \bibnamefont {Turro}}, \bibinfo {author}
  {\bibfnamefont {P.}~\bibnamefont {Kim}},\ and\ \bibinfo {author}
  {\bibfnamefont {L.}~\bibnamefont {Brus}},\ }\bibfield  {title} {\bibinfo
  {title} {Charge transfer chemical doping of few layer graphenes: Charge
  distribution and band gap formation},\ }\href
  {https://doi.org/10.1021/nl902362q} {\bibfield  {journal} {\bibinfo
  {journal} {Nano Letters}\ }\textbf {\bibinfo {volume} {9}},\ \bibinfo {pages}
  {4133} (\bibinfo {year} {2009})},\ \bibinfo {note} {pMID: 19827759},\ \Eprint
  {https://arxiv.org/abs/https://doi.org/10.1021/nl902362q}
  {https://doi.org/10.1021/nl902362q} \BibitemShut {NoStop}%
\bibitem [{\citenamefont {Crowther}\ \emph {et~al.}(2012)\citenamefont
  {Crowther}, \citenamefont {Ghassaei}, \citenamefont {Jung},\ and\
  \citenamefont {Brus}}]{Brusa}%
  \BibitemOpen
  \bibfield  {author} {\bibinfo {author} {\bibfnamefont {A.~C.}\ \bibnamefont
  {Crowther}}, \bibinfo {author} {\bibfnamefont {A.}~\bibnamefont {Ghassaei}},
  \bibinfo {author} {\bibfnamefont {N.}~\bibnamefont {Jung}},\ and\ \bibinfo
  {author} {\bibfnamefont {L.~E.}\ \bibnamefont {Brus}},\ }\bibfield  {title}
  {\bibinfo {title} {Strong charge-transfer doping of 1 to 10 layer graphene by
  no2},\ }\href {https://doi.org/10.1021/nn300252a} {\bibfield  {journal}
  {\bibinfo  {journal} {ACS Nano}\ }\textbf {\bibinfo {volume} {6}},\ \bibinfo
  {pages} {1865} (\bibinfo {year} {2012})},\ \bibinfo {note} {pMID: 22276666},\
  \Eprint {https://arxiv.org/abs/https://doi.org/10.1021/nn300252a}
  {https://doi.org/10.1021/nn300252a} \BibitemShut {NoStop}%
\bibitem [{\citenamefont {Bokdam}\ \emph {et~al.}(2011)\citenamefont {Bokdam},
  \citenamefont {Khomyakov}, \citenamefont {Brocks}, \citenamefont {Zhong},\
  and\ \citenamefont {Kelly}}]{KellyAndZhong}%
  \BibitemOpen
  \bibfield  {author} {\bibinfo {author} {\bibfnamefont {M.}~\bibnamefont
  {Bokdam}}, \bibinfo {author} {\bibfnamefont {P.~A.}\ \bibnamefont
  {Khomyakov}}, \bibinfo {author} {\bibfnamefont {G.}~\bibnamefont {Brocks}},
  \bibinfo {author} {\bibfnamefont {Z.}~\bibnamefont {Zhong}},\ and\ \bibinfo
  {author} {\bibfnamefont {P.~J.}\ \bibnamefont {Kelly}},\ }\bibfield  {title}
  {\bibinfo {title} {Electrostatic doping of graphene through ultrathin
  hexagonal boron nitride films},\ }\href {https://doi.org/10.1021/nl202131q}
  {\bibfield  {journal} {\bibinfo  {journal} {Nano Letters}\ }\textbf {\bibinfo
  {volume} {11}},\ \bibinfo {pages} {4631} (\bibinfo {year} {2011})},\ \bibinfo
  {note} {pMID: 21936569},\ \Eprint
  {https://arxiv.org/abs/https://doi.org/10.1021/nl202131q}
  {https://doi.org/10.1021/nl202131q} \BibitemShut {NoStop}%
\bibitem [{\citenamefont {Padilha}\ \emph {et~al.}(2015)\citenamefont
  {Padilha}, \citenamefont {Fazzio},\ and\ \citenamefont
  {da~Silva}}]{PhysRevLett.114.066803}%
  \BibitemOpen
  \bibfield  {author} {\bibinfo {author} {\bibfnamefont {J.~E.}\ \bibnamefont
  {Padilha}}, \bibinfo {author} {\bibfnamefont {A.}~\bibnamefont {Fazzio}},\
  and\ \bibinfo {author} {\bibfnamefont {A.~J.~R.}\ \bibnamefont {da~Silva}},\
  }\bibfield  {title} {\bibinfo {title} {van der waals heterostructure of
  phosphorene and graphene: Tuning the schottky barrier and doping by
  electrostatic gating},\ }\href
  {https://doi.org/10.1103/PhysRevLett.114.066803} {\bibfield  {journal}
  {\bibinfo  {journal} {Phys. Rev. Lett.}\ }\textbf {\bibinfo {volume} {114}},\
  \bibinfo {pages} {066803} (\bibinfo {year} {2015})}\BibitemShut {NoStop}%
\bibitem [{\citenamefont {Shi}\ \emph {et~al.}(2014)\citenamefont {Shi},
  \citenamefont {Tang}, \citenamefont {Zeng}, \citenamefont {Ju}, \citenamefont
  {Zhou}, \citenamefont {Zettl},\ and\ \citenamefont {Wang}}]{WangAndZettl}%
  \BibitemOpen
  \bibfield  {author} {\bibinfo {author} {\bibfnamefont {S.-F.}\ \bibnamefont
  {Shi}}, \bibinfo {author} {\bibfnamefont {T.-T.}\ \bibnamefont {Tang}},
  \bibinfo {author} {\bibfnamefont {B.}~\bibnamefont {Zeng}}, \bibinfo {author}
  {\bibfnamefont {L.}~\bibnamefont {Ju}}, \bibinfo {author} {\bibfnamefont
  {Q.}~\bibnamefont {Zhou}}, \bibinfo {author} {\bibfnamefont {A.}~\bibnamefont
  {Zettl}},\ and\ \bibinfo {author} {\bibfnamefont {F.}~\bibnamefont {Wang}},\
  }\bibfield  {title} {\bibinfo {title} {Controlling graphene ultrafast hot
  carrier response from metal-like to semiconductor-like by electrostatic
  gating},\ }\href {https://doi.org/10.1021/nl404826r} {\bibfield  {journal}
  {\bibinfo  {journal} {Nano Letters}\ }\textbf {\bibinfo {volume} {14}},\
  \bibinfo {pages} {1578} (\bibinfo {year} {2014})},\ \bibinfo {note} {pMID:
  24564302},\ \Eprint {https://arxiv.org/abs/https://doi.org/10.1021/nl404826r}
  {https://doi.org/10.1021/nl404826r} \BibitemShut {NoStop}%
\bibitem [{\citenamefont {Zhang}\ \emph
  {et~al.}(2009{\natexlab{a}})\citenamefont {Zhang}, \citenamefont {Tang},
  \citenamefont {Girit}, \citenamefont {Hao}, \citenamefont {Martin},
  \citenamefont {Zettl}, \citenamefont {Crommie}, \citenamefont {Shen},\ and\
  \citenamefont {Wang}}]{zhang_direct_2009}%
  \BibitemOpen
  \bibfield  {author} {\bibinfo {author} {\bibfnamefont {Y.}~\bibnamefont
  {Zhang}}, \bibinfo {author} {\bibfnamefont {T.-T.}\ \bibnamefont {Tang}},
  \bibinfo {author} {\bibfnamefont {C.}~\bibnamefont {Girit}}, \bibinfo
  {author} {\bibfnamefont {Z.}~\bibnamefont {Hao}}, \bibinfo {author}
  {\bibfnamefont {M.~C.}\ \bibnamefont {Martin}}, \bibinfo {author}
  {\bibfnamefont {A.}~\bibnamefont {Zettl}}, \bibinfo {author} {\bibfnamefont
  {M.~F.}\ \bibnamefont {Crommie}}, \bibinfo {author} {\bibfnamefont {Y.~R.}\
  \bibnamefont {Shen}},\ and\ \bibinfo {author} {\bibfnamefont
  {F.}~\bibnamefont {Wang}},\ }\bibfield  {title} {\bibinfo {title} {Direct
  observation of a widely tunable bandgap in bilayer graphene},\ }\href
  {https://doi.org/10.1038/nature08105} {\bibfield  {journal} {\bibinfo
  {journal} {Nature}\ }\textbf {\bibinfo {volume} {459}},\ \bibinfo {pages}
  {820} (\bibinfo {year} {2009}{\natexlab{a}})}\BibitemShut {NoStop}%
\bibitem [{\citenamefont {Bistritzer}\ and\ \citenamefont
  {MacDonald}(2011)}]{bistritzer_moire_2011}%
  \BibitemOpen
  \bibfield  {author} {\bibinfo {author} {\bibfnamefont {R.}~\bibnamefont
  {Bistritzer}}\ and\ \bibinfo {author} {\bibfnamefont {A.~H.}\ \bibnamefont
  {MacDonald}},\ }\bibfield  {title} {\bibinfo {title} {Moiré bands in twisted
  double-layer graphene},\ }\href {https://doi.org/10.1073/pnas.1108174108}
  {\bibfield  {journal} {\bibinfo  {journal} {Proceedings of the National
  Academy of Sciences of the United States of America}\ }\textbf {\bibinfo
  {volume} {108}},\ \bibinfo {pages} {12233} (\bibinfo {year} {2011})},\
  \bibinfo {note} {arXiv: 1009.4203 ISBN: 1108174108}\BibitemShut {NoStop}%
\bibitem [{\citenamefont {Wu}\ \emph {et~al.}(2018)\citenamefont {Wu},
  \citenamefont {Lovorn}, \citenamefont {Tutuc},\ and\ \citenamefont
  {MacDonald}}]{PhysRevLett.121.026402}%
  \BibitemOpen
  \bibfield  {author} {\bibinfo {author} {\bibfnamefont {F.}~\bibnamefont
  {Wu}}, \bibinfo {author} {\bibfnamefont {T.}~\bibnamefont {Lovorn}}, \bibinfo
  {author} {\bibfnamefont {E.}~\bibnamefont {Tutuc}},\ and\ \bibinfo {author}
  {\bibfnamefont {A.~H.}\ \bibnamefont {MacDonald}},\ }\bibfield  {title}
  {\bibinfo {title} {Hubbard model physics in transition metal dichalcogenide
  moir\'e bands},\ }\href {https://doi.org/10.1103/PhysRevLett.121.026402}
  {\bibfield  {journal} {\bibinfo  {journal} {Phys. Rev. Lett.}\ }\textbf
  {\bibinfo {volume} {121}},\ \bibinfo {pages} {026402} (\bibinfo {year}
  {2018})}\BibitemShut {NoStop}%
\bibitem [{\citenamefont {Honerkamp}(2008)}]{PhysRevLett.100.146404}%
  \BibitemOpen
  \bibfield  {author} {\bibinfo {author} {\bibfnamefont {C.}~\bibnamefont
  {Honerkamp}},\ }\bibfield  {title} {\bibinfo {title} {Density waves and
  cooper pairing on the honeycomb lattice},\ }\href
  {https://doi.org/10.1103/PhysRevLett.100.146404} {\bibfield  {journal}
  {\bibinfo  {journal} {Phys. Rev. Lett.}\ }\textbf {\bibinfo {volume} {100}},\
  \bibinfo {pages} {146404} (\bibinfo {year} {2008})}\BibitemShut {NoStop}%
\bibitem [{\citenamefont {Nandkishore}\ \emph {et~al.}(2012)\citenamefont
  {Nandkishore}, \citenamefont {Levitov},\ and\ \citenamefont
  {Chubukov}}]{nandkishore_chiral_2012}%
  \BibitemOpen
  \bibfield  {author} {\bibinfo {author} {\bibfnamefont {R.}~\bibnamefont
  {Nandkishore}}, \bibinfo {author} {\bibfnamefont {L.~S.}\ \bibnamefont
  {Levitov}},\ and\ \bibinfo {author} {\bibfnamefont {A.~V.}\ \bibnamefont
  {Chubukov}},\ }\bibfield  {title} {\bibinfo {title} {Chiral superconductivity
  from repulsive interactions in doped graphene},\ }\href
  {https://doi.org/10.1038/nphys2208} {\bibfield  {journal} {\bibinfo
  {journal} {Nature Physics}\ }\textbf {\bibinfo {volume} {8}},\ \bibinfo
  {pages} {158} (\bibinfo {year} {2012})}\BibitemShut {NoStop}%
\bibitem [{\citenamefont {Herbut}(2006)}]{PhysRevLett.97.146401}%
  \BibitemOpen
  \bibfield  {author} {\bibinfo {author} {\bibfnamefont {I.~F.}\ \bibnamefont
  {Herbut}},\ }\bibfield  {title} {\bibinfo {title} {Interactions and phase
  transitions on graphene's honeycomb lattice},\ }\href
  {https://doi.org/10.1103/PhysRevLett.97.146401} {\bibfield  {journal}
  {\bibinfo  {journal} {Phys. Rev. Lett.}\ }\textbf {\bibinfo {volume} {97}},\
  \bibinfo {pages} {146401} (\bibinfo {year} {2006})}\BibitemShut {NoStop}%
\bibitem [{\citenamefont {Song}\ \emph {et~al.}(2012)\citenamefont {Song},
  \citenamefont {Sun}, \citenamefont {Wang}, \citenamefont {Jiang},
  \citenamefont {Wang}, \citenamefont {He}, \citenamefont {Chen}, \citenamefont
  {Zhang}, \citenamefont {Ma},\ and\ \citenamefont
  {Xue}}]{PhysRevLett.108.156803}%
  \BibitemOpen
  \bibfield  {author} {\bibinfo {author} {\bibfnamefont {C.-L.}\ \bibnamefont
  {Song}}, \bibinfo {author} {\bibfnamefont {B.}~\bibnamefont {Sun}}, \bibinfo
  {author} {\bibfnamefont {Y.-L.}\ \bibnamefont {Wang}}, \bibinfo {author}
  {\bibfnamefont {Y.-P.}\ \bibnamefont {Jiang}}, \bibinfo {author}
  {\bibfnamefont {L.}~\bibnamefont {Wang}}, \bibinfo {author} {\bibfnamefont
  {K.}~\bibnamefont {He}}, \bibinfo {author} {\bibfnamefont {X.}~\bibnamefont
  {Chen}}, \bibinfo {author} {\bibfnamefont {P.}~\bibnamefont {Zhang}},
  \bibinfo {author} {\bibfnamefont {X.-C.}\ \bibnamefont {Ma}},\ and\ \bibinfo
  {author} {\bibfnamefont {Q.-K.}\ \bibnamefont {Xue}},\ }\bibfield  {title}
  {\bibinfo {title} {Charge-transfer-induced cesium superlattices on
  graphene},\ }\href {https://doi.org/10.1103/PhysRevLett.108.156803}
  {\bibfield  {journal} {\bibinfo  {journal} {Phys. Rev. Lett.}\ }\textbf
  {\bibinfo {volume} {108}},\ \bibinfo {pages} {156803} (\bibinfo {year}
  {2012})}\BibitemShut {NoStop}%
\bibitem [{\citenamefont {Efetov}\ \emph {et~al.}(2011)\citenamefont {Efetov},
  \citenamefont {Maher}, \citenamefont {Glinskis},\ and\ \citenamefont
  {Kim}}]{PhysRevB.84.161412}%
  \BibitemOpen
  \bibfield  {author} {\bibinfo {author} {\bibfnamefont {D.~K.}\ \bibnamefont
  {Efetov}}, \bibinfo {author} {\bibfnamefont {P.}~\bibnamefont {Maher}},
  \bibinfo {author} {\bibfnamefont {S.}~\bibnamefont {Glinskis}},\ and\
  \bibinfo {author} {\bibfnamefont {P.}~\bibnamefont {Kim}},\ }\bibfield
  {title} {\bibinfo {title} {Multiband transport in bilayer graphene at high
  carrier densities},\ }\href {https://doi.org/10.1103/PhysRevB.84.161412}
  {\bibfield  {journal} {\bibinfo  {journal} {Phys. Rev. B}\ }\textbf {\bibinfo
  {volume} {84}},\ \bibinfo {pages} {161412} (\bibinfo {year}
  {2011})}\BibitemShut {NoStop}%
\bibitem [{\citenamefont {Zhao}\ \emph {et~al.}(2011)\citenamefont {Zhao},
  \citenamefont {He}, \citenamefont {Rim}, \citenamefont {Schiros},
  \citenamefont {Kim}, \citenamefont {Zhou}, \citenamefont {Gutiérrez},
  \citenamefont {Chockalingam}, \citenamefont {Arguello}, \citenamefont
  {Pálová}, \citenamefont {Nordlund}, \citenamefont {Hybertsen},
  \citenamefont {Reichman}, \citenamefont {Heinz}, \citenamefont {Kim},
  \citenamefont {Pinczuk}, \citenamefont {Flynn},\ and\ \citenamefont
  {Pasupathy}}]{PasupathyScience}%
  \BibitemOpen
  \bibfield  {author} {\bibinfo {author} {\bibfnamefont {L.}~\bibnamefont
  {Zhao}}, \bibinfo {author} {\bibfnamefont {R.}~\bibnamefont {He}}, \bibinfo
  {author} {\bibfnamefont {K.~T.}\ \bibnamefont {Rim}}, \bibinfo {author}
  {\bibfnamefont {T.}~\bibnamefont {Schiros}}, \bibinfo {author} {\bibfnamefont
  {K.~S.}\ \bibnamefont {Kim}}, \bibinfo {author} {\bibfnamefont
  {H.}~\bibnamefont {Zhou}}, \bibinfo {author} {\bibfnamefont {C.}~\bibnamefont
  {Gutiérrez}}, \bibinfo {author} {\bibfnamefont {S.~P.}\ \bibnamefont
  {Chockalingam}}, \bibinfo {author} {\bibfnamefont {C.~J.}\ \bibnamefont
  {Arguello}}, \bibinfo {author} {\bibfnamefont {L.}~\bibnamefont {Pálová}},
  \bibinfo {author} {\bibfnamefont {D.}~\bibnamefont {Nordlund}}, \bibinfo
  {author} {\bibfnamefont {M.~S.}\ \bibnamefont {Hybertsen}}, \bibinfo {author}
  {\bibfnamefont {D.~R.}\ \bibnamefont {Reichman}}, \bibinfo {author}
  {\bibfnamefont {T.~F.}\ \bibnamefont {Heinz}}, \bibinfo {author}
  {\bibfnamefont {P.}~\bibnamefont {Kim}}, \bibinfo {author} {\bibfnamefont
  {A.}~\bibnamefont {Pinczuk}}, \bibinfo {author} {\bibfnamefont {G.~W.}\
  \bibnamefont {Flynn}},\ and\ \bibinfo {author} {\bibfnamefont {A.~N.}\
  \bibnamefont {Pasupathy}},\ }\bibfield  {title} {\bibinfo {title}
  {Visualizing individual nitrogen dopants in monolayer graphene},\ }\href
  {https://doi.org/10.1126/science.1208759} {\bibfield  {journal} {\bibinfo
  {journal} {Science}\ }\textbf {\bibinfo {volume} {333}},\ \bibinfo {pages}
  {999} (\bibinfo {year} {2011})},\ \Eprint
  {https://arxiv.org/abs/https://www.science.org/doi/pdf/10.1126/science.1208759}
  {https://www.science.org/doi/pdf/10.1126/science.1208759} \BibitemShut
  {NoStop}%
\bibitem [{\citenamefont {Kim}\ \emph {et~al.}(2016)\citenamefont {Kim},
  \citenamefont {Dugerjav}, \citenamefont {Lkhagvasuren},\ and\ \citenamefont
  {Seo}}]{Kim_2016}%
  \BibitemOpen
  \bibfield  {author} {\bibinfo {author} {\bibfnamefont {H.}~\bibnamefont
  {Kim}}, \bibinfo {author} {\bibfnamefont {O.}~\bibnamefont {Dugerjav}},
  \bibinfo {author} {\bibfnamefont {A.}~\bibnamefont {Lkhagvasuren}},\ and\
  \bibinfo {author} {\bibfnamefont {J.~M.}\ \bibnamefont {Seo}},\ }\bibfield
  {title} {\bibinfo {title} {Charge neutrality of quasi-free-standing monolayer
  graphene induced by the intercalated sn layer},\ }\href
  {https://doi.org/10.1088/0022-3727/49/13/135307} {\bibfield  {journal}
  {\bibinfo  {journal} {Journal of Physics D: Applied Physics}\ }\textbf
  {\bibinfo {volume} {49}},\ \bibinfo {pages} {135307} (\bibinfo {year}
  {2016})}\BibitemShut {NoStop}%
\bibitem [{\citenamefont {Yurtsever}\ \emph {et~al.}(2016)\citenamefont
  {Yurtsever}, \citenamefont {Onoda}, \citenamefont {Iimori}, \citenamefont
  {Niki}, \citenamefont {Miyamachi}, \citenamefont {Abe}, \citenamefont
  {Mizuno}, \citenamefont {Tanaka}, \citenamefont {Komori},\ and\ \citenamefont
  {Sugimoto}}]{yurtsever_effects_2016}%
  \BibitemOpen
  \bibfield  {author} {\bibinfo {author} {\bibfnamefont {A.}~\bibnamefont
  {Yurtsever}}, \bibinfo {author} {\bibfnamefont {J.}~\bibnamefont {Onoda}},
  \bibinfo {author} {\bibfnamefont {T.}~\bibnamefont {Iimori}}, \bibinfo
  {author} {\bibfnamefont {K.}~\bibnamefont {Niki}}, \bibinfo {author}
  {\bibfnamefont {T.}~\bibnamefont {Miyamachi}}, \bibinfo {author}
  {\bibfnamefont {M.}~\bibnamefont {Abe}}, \bibinfo {author} {\bibfnamefont
  {S.}~\bibnamefont {Mizuno}}, \bibinfo {author} {\bibfnamefont
  {S.}~\bibnamefont {Tanaka}}, \bibinfo {author} {\bibfnamefont
  {F.}~\bibnamefont {Komori}},\ and\ \bibinfo {author} {\bibfnamefont
  {Y.}~\bibnamefont {Sugimoto}},\ }\bibfield  {title} {\bibinfo {title}
  {Effects of {Pb} {Intercalation} on the {Structural} and {Electronic}
  {Properties} of {Epitaxial} {Graphene} on {SiC}},\ }\href
  {https://doi.org/10.1002/smll.201600666} {\bibfield  {journal} {\bibinfo
  {journal} {Small}\ }\textbf {\bibinfo {volume} {12}},\ \bibinfo {pages}
  {3956} (\bibinfo {year} {2016})}\BibitemShut {NoStop}%
\bibitem [{\citenamefont {Briggs}\ \emph {et~al.}(2019)\citenamefont {Briggs},
  \citenamefont {Gebeyehu}, \citenamefont {Vera}, \citenamefont {Zhao},
  \citenamefont {Wang}, \citenamefont {De~La Fuente~Duran}, \citenamefont
  {Bersch}, \citenamefont {Bowen}, \citenamefont {Knappenberger},\ and\
  \citenamefont {Robinson}}]{briggs_epitaxial_2019}%
  \BibitemOpen
  \bibfield  {author} {\bibinfo {author} {\bibfnamefont {N.}~\bibnamefont
  {Briggs}}, \bibinfo {author} {\bibfnamefont {Z.~M.}\ \bibnamefont
  {Gebeyehu}}, \bibinfo {author} {\bibfnamefont {A.}~\bibnamefont {Vera}},
  \bibinfo {author} {\bibfnamefont {T.}~\bibnamefont {Zhao}}, \bibinfo {author}
  {\bibfnamefont {K.}~\bibnamefont {Wang}}, \bibinfo {author} {\bibfnamefont
  {A.}~\bibnamefont {De~La Fuente~Duran}}, \bibinfo {author} {\bibfnamefont
  {B.}~\bibnamefont {Bersch}}, \bibinfo {author} {\bibfnamefont
  {T.}~\bibnamefont {Bowen}}, \bibinfo {author} {\bibfnamefont {K.~L.}\
  \bibnamefont {Knappenberger}},\ and\ \bibinfo {author} {\bibfnamefont
  {J.~A.}\ \bibnamefont {Robinson}},\ }\bibfield  {title} {\bibinfo {title}
  {Epitaxial graphene/silicon carbide intercalation: a minireview on graphene
  modulation and unique {2D} materials},\ }\href
  {https://doi.org/10.1039/C9NR03721G} {\bibfield  {journal} {\bibinfo
  {journal} {Nanoscale}\ }\textbf {\bibinfo {volume} {11}},\ \bibinfo {pages}
  {15440} (\bibinfo {year} {2019})}\BibitemShut {NoStop}%
\bibitem [{\citenamefont {Rosenzweig}\ \emph {et~al.}(2019)\citenamefont
  {Rosenzweig}, \citenamefont {Karakachian}, \citenamefont {Link},
  \citenamefont {K\"uster},\ and\ \citenamefont {Starke}}]{RosenzweigPRB}%
  \BibitemOpen
  \bibfield  {author} {\bibinfo {author} {\bibfnamefont {P.}~\bibnamefont
  {Rosenzweig}}, \bibinfo {author} {\bibfnamefont {H.}~\bibnamefont
  {Karakachian}}, \bibinfo {author} {\bibfnamefont {S.}~\bibnamefont {Link}},
  \bibinfo {author} {\bibfnamefont {K.}~\bibnamefont {K\"uster}},\ and\
  \bibinfo {author} {\bibfnamefont {U.}~\bibnamefont {Starke}},\ }\bibfield
  {title} {\bibinfo {title} {Tuning the doping level of graphene in the
  vicinity of the van hove singularity via ytterbium intercalation},\ }\href
  {https://doi.org/10.1103/PhysRevB.100.035445} {\bibfield  {journal} {\bibinfo
   {journal} {Phys. Rev. B}\ }\textbf {\bibinfo {volume} {100}},\ \bibinfo
  {pages} {035445} (\bibinfo {year} {2019})}\BibitemShut {NoStop}%
\bibitem [{\citenamefont {Rosenzweig}\ \emph {et~al.}(2020)\citenamefont
  {Rosenzweig}, \citenamefont {Karakachian}, \citenamefont {Marchenko},
  \citenamefont {K\"uster},\ and\ \citenamefont {Starke}}]{RosenzweigPRL}%
  \BibitemOpen
  \bibfield  {author} {\bibinfo {author} {\bibfnamefont {P.}~\bibnamefont
  {Rosenzweig}}, \bibinfo {author} {\bibfnamefont {H.}~\bibnamefont
  {Karakachian}}, \bibinfo {author} {\bibfnamefont {D.}~\bibnamefont
  {Marchenko}}, \bibinfo {author} {\bibfnamefont {K.}~\bibnamefont
  {K\"uster}},\ and\ \bibinfo {author} {\bibfnamefont {U.}~\bibnamefont
  {Starke}},\ }\bibfield  {title} {\bibinfo {title} {Overdoping graphene beyond
  the van hove singularity},\ }\href
  {https://doi.org/10.1103/PhysRevLett.125.176403} {\bibfield  {journal}
  {\bibinfo  {journal} {Phys. Rev. Lett.}\ }\textbf {\bibinfo {volume} {125}},\
  \bibinfo {pages} {176403} (\bibinfo {year} {2020})}\BibitemShut {NoStop}%
\bibitem [{\citenamefont {Pašti}\ \emph {et~al.}(2018)\citenamefont {Pašti},
  \citenamefont {Jovanović}, \citenamefont {Dobrota}, \citenamefont {Mentus},
  \citenamefont {Johansson},\ and\ \citenamefont
  {Skorodumova}}]{pasti_atomic_2018}%
  \BibitemOpen
  \bibfield  {author} {\bibinfo {author} {\bibfnamefont {I.~A.}\ \bibnamefont
  {Pašti}}, \bibinfo {author} {\bibfnamefont {A.}~\bibnamefont {Jovanović}},
  \bibinfo {author} {\bibfnamefont {A.~S.}\ \bibnamefont {Dobrota}}, \bibinfo
  {author} {\bibfnamefont {S.~V.}\ \bibnamefont {Mentus}}, \bibinfo {author}
  {\bibfnamefont {B.}~\bibnamefont {Johansson}},\ and\ \bibinfo {author}
  {\bibfnamefont {N.~V.}\ \bibnamefont {Skorodumova}},\ }\bibfield  {title}
  {\bibinfo {title} {Atomic adsorption on pristine graphene along the
  {Periodic} {Table} of {Elements} – {From} {PBE} to non-local functionals},\
  }\href {https://doi.org/https://doi.org/10.1016/j.apsusc.2017.12.046}
  {\bibfield  {journal} {\bibinfo  {journal} {Applied Surface Science}\
  }\textbf {\bibinfo {volume} {436}},\ \bibinfo {pages} {433} (\bibinfo {year}
  {2018})}\BibitemShut {NoStop}%
\bibitem [{\citenamefont {Kim}\ \emph {et~al.}(2017)\citenamefont {Kim},
  \citenamefont {Tringides}, \citenamefont {Hershberger}, \citenamefont {Chen},
  \citenamefont {Hupalo}, \citenamefont {Thiel}, \citenamefont {Wang},\ and\
  \citenamefont {Ho}}]{kim_manipulation_2017}%
  \BibitemOpen
  \bibfield  {author} {\bibinfo {author} {\bibfnamefont {M.}~\bibnamefont
  {Kim}}, \bibinfo {author} {\bibfnamefont {M.~C.}\ \bibnamefont {Tringides}},
  \bibinfo {author} {\bibfnamefont {M.~T.}\ \bibnamefont {Hershberger}},
  \bibinfo {author} {\bibfnamefont {S.}~\bibnamefont {Chen}}, \bibinfo {author}
  {\bibfnamefont {M.}~\bibnamefont {Hupalo}}, \bibinfo {author} {\bibfnamefont
  {P.~A.}\ \bibnamefont {Thiel}}, \bibinfo {author} {\bibfnamefont {C.-Z.}\
  \bibnamefont {Wang}},\ and\ \bibinfo {author} {\bibfnamefont {K.-M.}\
  \bibnamefont {Ho}},\ }\bibfield  {title} {\bibinfo {title} {Manipulation of
  {Dirac} cones in intercalated epitaxial graphene},\ }\href
  {https://doi.org/https://doi.org/10.1016/j.carbon.2017.07.020} {\bibfield
  {journal} {\bibinfo  {journal} {Carbon}\ }\textbf {\bibinfo {volume} {123}},\
  \bibinfo {pages} {93} (\bibinfo {year} {2017})}\BibitemShut {NoStop}%
\bibitem [{\citenamefont {Daukiya}\ \emph {et~al.}(2018)\citenamefont
  {Daukiya}, \citenamefont {Nair}, \citenamefont {Hajjar-Garreau},
  \citenamefont {Vonau}, \citenamefont {Aubel}, \citenamefont {Bubendorff},
  \citenamefont {Cranney}, \citenamefont {Denys}, \citenamefont {Florentin},
  \citenamefont {Reiter},\ and\ \citenamefont {Simon}}]{SimonPhysRevB}%
  \BibitemOpen
  \bibfield  {author} {\bibinfo {author} {\bibfnamefont {L.}~\bibnamefont
  {Daukiya}}, \bibinfo {author} {\bibfnamefont {M.~N.}\ \bibnamefont {Nair}},
  \bibinfo {author} {\bibfnamefont {S.}~\bibnamefont {Hajjar-Garreau}},
  \bibinfo {author} {\bibfnamefont {F.}~\bibnamefont {Vonau}}, \bibinfo
  {author} {\bibfnamefont {D.}~\bibnamefont {Aubel}}, \bibinfo {author}
  {\bibfnamefont {J.~L.}\ \bibnamefont {Bubendorff}}, \bibinfo {author}
  {\bibfnamefont {M.}~\bibnamefont {Cranney}}, \bibinfo {author} {\bibfnamefont
  {E.}~\bibnamefont {Denys}}, \bibinfo {author} {\bibfnamefont
  {A.}~\bibnamefont {Florentin}}, \bibinfo {author} {\bibfnamefont
  {G.}~\bibnamefont {Reiter}},\ and\ \bibinfo {author} {\bibfnamefont
  {L.}~\bibnamefont {Simon}},\ }\bibfield  {title} {\bibinfo {title} {Highly
  $n$-doped graphene generated through intercalated terbium atoms},\ }\href
  {https://doi.org/10.1103/PhysRevB.97.035309} {\bibfield  {journal} {\bibinfo
  {journal} {Phys. Rev. B}\ }\textbf {\bibinfo {volume} {97}},\ \bibinfo
  {pages} {035309} (\bibinfo {year} {2018})}\BibitemShut {NoStop}%
\bibitem [{\citenamefont {Irkhin}\ \emph {et~al.}(2002)\citenamefont {Irkhin},
  \citenamefont {Katanin},\ and\ \citenamefont
  {Katsnelson}}]{IrkhinPhysRevLett}%
  \BibitemOpen
  \bibfield  {author} {\bibinfo {author} {\bibfnamefont {V.~Y.}\ \bibnamefont
  {Irkhin}}, \bibinfo {author} {\bibfnamefont {A.~A.}\ \bibnamefont
  {Katanin}},\ and\ \bibinfo {author} {\bibfnamefont {M.~I.}\ \bibnamefont
  {Katsnelson}},\ }\bibfield  {title} {\bibinfo {title} {Robustness of the van
  hove scenario for high-${T}_{c}$ superconductors},\ }\href
  {https://doi.org/10.1103/PhysRevLett.89.076401} {\bibfield  {journal}
  {\bibinfo  {journal} {Phys. Rev. Lett.}\ }\textbf {\bibinfo {volume} {89}},\
  \bibinfo {pages} {076401} (\bibinfo {year} {2002})}\BibitemShut {NoStop}%
\bibitem [{\citenamefont {Yudin}\ \emph {et~al.}(2014)\citenamefont {Yudin},
  \citenamefont {Hirschmeier}, \citenamefont {Hafermann}, \citenamefont
  {Eriksson}, \citenamefont {Lichtenstein},\ and\ \citenamefont
  {Katsnelson}}]{YudinPhysRevLett}%
  \BibitemOpen
  \bibfield  {author} {\bibinfo {author} {\bibfnamefont {D.}~\bibnamefont
  {Yudin}}, \bibinfo {author} {\bibfnamefont {D.}~\bibnamefont {Hirschmeier}},
  \bibinfo {author} {\bibfnamefont {H.}~\bibnamefont {Hafermann}}, \bibinfo
  {author} {\bibfnamefont {O.}~\bibnamefont {Eriksson}}, \bibinfo {author}
  {\bibfnamefont {A.~I.}\ \bibnamefont {Lichtenstein}},\ and\ \bibinfo {author}
  {\bibfnamefont {M.~I.}\ \bibnamefont {Katsnelson}},\ }\bibfield  {title}
  {\bibinfo {title} {Fermi condensation near van hove singularities within the
  hubbard model on the triangular lattice},\ }\href
  {https://doi.org/10.1103/PhysRevLett.112.070403} {\bibfield  {journal}
  {\bibinfo  {journal} {Phys. Rev. Lett.}\ }\textbf {\bibinfo {volume} {112}},\
  \bibinfo {pages} {070403} (\bibinfo {year} {2014})}\BibitemShut {NoStop}%
\bibitem [{\citenamefont {Zaarour}\ \emph {et~al.}(2023)\citenamefont
  {Zaarour}, \citenamefont {Malesys}, \citenamefont {Teyssandier},
  \citenamefont {Cranney}, \citenamefont {Denys}, \citenamefont {Bubendorff},
  \citenamefont {Florentin}, \citenamefont {Josien}, \citenamefont {Vonau},
  \citenamefont {Aubel}, \citenamefont {Ouerghi}, \citenamefont {Bena},\ and\
  \citenamefont {Simon}}]{ZaarourPhysRevResearch}%
  \BibitemOpen
  \bibfield  {author} {\bibinfo {author} {\bibfnamefont {A.}~\bibnamefont
  {Zaarour}}, \bibinfo {author} {\bibfnamefont {V.}~\bibnamefont {Malesys}},
  \bibinfo {author} {\bibfnamefont {J.}~\bibnamefont {Teyssandier}}, \bibinfo
  {author} {\bibfnamefont {M.}~\bibnamefont {Cranney}}, \bibinfo {author}
  {\bibfnamefont {E.}~\bibnamefont {Denys}}, \bibinfo {author} {\bibfnamefont
  {J.~L.}\ \bibnamefont {Bubendorff}}, \bibinfo {author} {\bibfnamefont
  {A.}~\bibnamefont {Florentin}}, \bibinfo {author} {\bibfnamefont
  {L.}~\bibnamefont {Josien}}, \bibinfo {author} {\bibfnamefont
  {F.}~\bibnamefont {Vonau}}, \bibinfo {author} {\bibfnamefont
  {D.}~\bibnamefont {Aubel}}, \bibinfo {author} {\bibfnamefont
  {A.}~\bibnamefont {Ouerghi}}, \bibinfo {author} {\bibfnamefont
  {C.}~\bibnamefont {Bena}},\ and\ \bibinfo {author} {\bibfnamefont
  {L.}~\bibnamefont {Simon}},\ }\bibfield  {title} {\bibinfo {title} {Flat band
  and lifshitz transition in long-range-ordered supergraphene obtained by
  erbium intercalation},\ }\href
  {https://doi.org/10.1103/PhysRevResearch.5.013099} {\bibfield  {journal}
  {\bibinfo  {journal} {Phys. Rev. Res.}\ }\textbf {\bibinfo {volume} {5}},\
  \bibinfo {pages} {013099} (\bibinfo {year} {2023})}\BibitemShut {NoStop}%
\bibitem [{\citenamefont {Zhang}\ \emph
  {et~al.}(2009{\natexlab{b}})\citenamefont {Zhang}, \citenamefont {Brar},
  \citenamefont {Girit}, \citenamefont {Zettl},\ and\ \citenamefont
  {Crommie}}]{ZhangNaturePhys}%
  \BibitemOpen
  \bibfield  {author} {\bibinfo {author} {\bibfnamefont {Y.}~\bibnamefont
  {Zhang}}, \bibinfo {author} {\bibfnamefont {V.~W.}\ \bibnamefont {Brar}},
  \bibinfo {author} {\bibfnamefont {C.}~\bibnamefont {Girit}}, \bibinfo
  {author} {\bibfnamefont {A.}~\bibnamefont {Zettl}},\ and\ \bibinfo {author}
  {\bibfnamefont {M.~F.}\ \bibnamefont {Crommie}},\ }\bibfield  {title}
  {\bibinfo {title} {Origin of spatial charge inhomogeneity in graphene},\
  }\href {https://doi.org/10.1038/nphys1365} {\bibfield  {journal} {\bibinfo
  {journal} {Nature Physics}\ }\textbf {\bibinfo {volume} {5}},\ \bibinfo
  {pages} {722} (\bibinfo {year} {2009}{\natexlab{b}})}\BibitemShut {NoStop}%
\bibitem [{\citenamefont {Martin}\ \emph {et~al.}(2008)\citenamefont {Martin},
  \citenamefont {Akerman}, \citenamefont {Ulbricht}, \citenamefont {Lohmann},
  \citenamefont {Smet}, \citenamefont {von Klitzing},\ and\ \citenamefont
  {Yacoby}}]{martin_observation_2008}%
  \BibitemOpen
  \bibfield  {author} {\bibinfo {author} {\bibfnamefont {J.}~\bibnamefont
  {Martin}}, \bibinfo {author} {\bibfnamefont {N.}~\bibnamefont {Akerman}},
  \bibinfo {author} {\bibfnamefont {G.}~\bibnamefont {Ulbricht}}, \bibinfo
  {author} {\bibfnamefont {T.}~\bibnamefont {Lohmann}}, \bibinfo {author}
  {\bibfnamefont {J.~H.}\ \bibnamefont {Smet}}, \bibinfo {author}
  {\bibfnamefont {K.}~\bibnamefont {von Klitzing}},\ and\ \bibinfo {author}
  {\bibfnamefont {A.}~\bibnamefont {Yacoby}},\ }\bibfield  {title} {\bibinfo
  {title} {Observation of electron–hole puddles in graphene using a scanning
  single-electron transistor},\ }\href {https://doi.org/10.1038/nphys781}
  {\bibfield  {journal} {\bibinfo  {journal} {Nature Physics}\ }\textbf
  {\bibinfo {volume} {4}},\ \bibinfo {pages} {144} (\bibinfo {year}
  {2008})}\BibitemShut {NoStop}%
\bibitem [{\citenamefont {Rutter}\ \emph {et~al.}(2007)\citenamefont {Rutter},
  \citenamefont {Crain}, \citenamefont {Guisinger}, \citenamefont {Li},
  \citenamefont {First},\ and\ \citenamefont {Stroscio}}]{Stroscio}%
  \BibitemOpen
  \bibfield  {author} {\bibinfo {author} {\bibfnamefont {G.~M.}\ \bibnamefont
  {Rutter}}, \bibinfo {author} {\bibfnamefont {J.~N.}\ \bibnamefont {Crain}},
  \bibinfo {author} {\bibfnamefont {N.~P.}\ \bibnamefont {Guisinger}}, \bibinfo
  {author} {\bibfnamefont {T.}~\bibnamefont {Li}}, \bibinfo {author}
  {\bibfnamefont {P.~N.}\ \bibnamefont {First}},\ and\ \bibinfo {author}
  {\bibfnamefont {J.~A.}\ \bibnamefont {Stroscio}},\ }\bibfield  {title}
  {\bibinfo {title} {Scattering and interference in epitaxial graphene},\
  }\href {https://doi.org/10.1126/science.1142882} {\bibfield  {journal}
  {\bibinfo  {journal} {Science}\ }\textbf {\bibinfo {volume} {317}},\ \bibinfo
  {pages} {219} (\bibinfo {year} {2007})},\ \Eprint
  {https://arxiv.org/abs/https://www.science.org/doi/pdf/10.1126/science.1142882}
  {https://www.science.org/doi/pdf/10.1126/science.1142882} \BibitemShut
  {NoStop}%
\bibitem [{\citenamefont {Crook}\ \emph {et~al.}(2015)\citenamefont {Crook},
  \citenamefont {Constantin}, \citenamefont {Ahmed}, \citenamefont {Zhu},
  \citenamefont {Balatsky},\ and\ \citenamefont
  {Haraldsen}}]{crook_proximity-induced_2015}%
  \BibitemOpen
  \bibfield  {author} {\bibinfo {author} {\bibfnamefont {C.~B.}\ \bibnamefont
  {Crook}}, \bibinfo {author} {\bibfnamefont {C.}~\bibnamefont {Constantin}},
  \bibinfo {author} {\bibfnamefont {T.}~\bibnamefont {Ahmed}}, \bibinfo
  {author} {\bibfnamefont {J.-X.}\ \bibnamefont {Zhu}}, \bibinfo {author}
  {\bibfnamefont {A.~V.}\ \bibnamefont {Balatsky}},\ and\ \bibinfo {author}
  {\bibfnamefont {J.~T.}\ \bibnamefont {Haraldsen}},\ }\bibfield  {title}
  {\bibinfo {title} {Proximity-induced magnetism in transition-metal
  substituted graphene},\ }\href {https://doi.org/10.1038/srep12322} {\bibfield
   {journal} {\bibinfo  {journal} {Scientific Reports}\ }\textbf {\bibinfo
  {volume} {5}},\ \bibinfo {pages} {12322} (\bibinfo {year}
  {2015})}\BibitemShut {NoStop}%
\bibitem [{\citenamefont {Basiuk}\ \emph {et~al.}(2022)\citenamefont {Basiuk},
  \citenamefont {Prezhdo},\ and\ \citenamefont
  {Basiuk}}]{basiuk_adsorption_2022}%
  \BibitemOpen
  \bibfield  {author} {\bibinfo {author} {\bibfnamefont {V.~A.}\ \bibnamefont
  {Basiuk}}, \bibinfo {author} {\bibfnamefont {O.~V.}\ \bibnamefont
  {Prezhdo}},\ and\ \bibinfo {author} {\bibfnamefont {E.~V.}\ \bibnamefont
  {Basiuk}},\ }\bibfield  {title} {\bibinfo {title} {Adsorption of {Lanthanide}
  {Atoms} on {Graphene}: {Similar}, {Yet} {Different}},\ }\href
  {https://doi.org/10.1021/acs.jpclett.2c01580} {\bibfield  {journal} {\bibinfo
   {journal} {The Journal of Physical Chemistry Letters}\ }\textbf {\bibinfo
  {volume} {13}},\ \bibinfo {pages} {6042} (\bibinfo {year} {2022})},\ \bibinfo
  {note} {publisher: American Chemical Society}\BibitemShut {NoStop}%
\bibitem [{\citenamefont {Hwang}\ \emph {et~al.}(2014)\citenamefont {Hwang},
  \citenamefont {Kim}, \citenamefont {Siegel}, \citenamefont {Chan},
  \citenamefont {Noffsinger}, \citenamefont {Fedorov}, \citenamefont {Cohen},
  \citenamefont {Johansson}, \citenamefont {Neaton},\ and\ \citenamefont
  {Lanzara}}]{HwangPRB}%
  \BibitemOpen
  \bibfield  {author} {\bibinfo {author} {\bibfnamefont {C.}~\bibnamefont
  {Hwang}}, \bibinfo {author} {\bibfnamefont {D.~Y.}\ \bibnamefont {Kim}},
  \bibinfo {author} {\bibfnamefont {D.~A.}\ \bibnamefont {Siegel}}, \bibinfo
  {author} {\bibfnamefont {K.~T.}\ \bibnamefont {Chan}}, \bibinfo {author}
  {\bibfnamefont {J.}~\bibnamefont {Noffsinger}}, \bibinfo {author}
  {\bibfnamefont {A.~V.}\ \bibnamefont {Fedorov}}, \bibinfo {author}
  {\bibfnamefont {M.~L.}\ \bibnamefont {Cohen}}, \bibinfo {author}
  {\bibfnamefont {B.}~\bibnamefont {Johansson}}, \bibinfo {author}
  {\bibfnamefont {J.~B.}\ \bibnamefont {Neaton}},\ and\ \bibinfo {author}
  {\bibfnamefont {A.}~\bibnamefont {Lanzara}},\ }\bibfield  {title} {\bibinfo
  {title} {Ytterbium-driven strong enhancement of electron-phonon coupling in
  graphene},\ }\href {https://doi.org/10.1103/PhysRevB.90.115417} {\bibfield
  {journal} {\bibinfo  {journal} {Phys. Rev. B}\ }\textbf {\bibinfo {volume}
  {90}},\ \bibinfo {pages} {115417} (\bibinfo {year} {2014})}\BibitemShut
  {NoStop}%
\bibitem [{\citenamefont {Kang}\ \emph {et~al.}(2019)\citenamefont {Kang},
  \citenamefont {Hwang}, \citenamefont {Lee}, \citenamefont {Fedorov},\ and\
  \citenamefont {Hwang}}]{KANG20191}%
  \BibitemOpen
  \bibfield  {author} {\bibinfo {author} {\bibfnamefont {M.}~\bibnamefont
  {Kang}}, \bibinfo {author} {\bibfnamefont {J.}~\bibnamefont {Hwang}},
  \bibinfo {author} {\bibfnamefont {J.-E.}\ \bibnamefont {Lee}}, \bibinfo
  {author} {\bibfnamefont {A.}~\bibnamefont {Fedorov}},\ and\ \bibinfo {author}
  {\bibfnamefont {C.}~\bibnamefont {Hwang}},\ }\bibfield  {title} {\bibinfo
  {title} {Concomitant enhancement of electron-phonon coupling and
  electron-electron interaction in graphene decorated with ytterbium},\ }\href
  {https://doi.org/https://doi.org/10.1016/j.apsusc.2018.10.134} {\bibfield
  {journal} {\bibinfo  {journal} {Applied Surface Science}\ }\textbf {\bibinfo
  {volume} {467-468}},\ \bibinfo {pages} {1} (\bibinfo {year}
  {2019})}\BibitemShut {NoStop}%
\bibitem [{\citenamefont {Kim}\ \emph {et~al.}(2022)\citenamefont {Kim},
  \citenamefont {Drozdov}, \citenamefont {Fujita}, \citenamefont {Davis},
  \citenamefont {Božović},\ and\ \citenamefont {Valla}}]{CKKim2018}%
  \BibitemOpen
  \bibfield  {author} {\bibinfo {author} {\bibfnamefont {C.~K.}\ \bibnamefont
  {Kim}}, \bibinfo {author} {\bibfnamefont {I.~K.}\ \bibnamefont {Drozdov}},
  \bibinfo {author} {\bibfnamefont {K.}~\bibnamefont {Fujita}}, \bibinfo
  {author} {\bibfnamefont {J.~C.~S.}\ \bibnamefont {Davis}}, \bibinfo {author}
  {\bibfnamefont {B.}~\bibnamefont {Božović}},\ and\ \bibinfo {author}
  {\bibfnamefont {T.}~\bibnamefont {Valla}},\ }\bibfield  {title} {\bibinfo
  {title} {In-situ angle-resolved photoemission spectroscopy of copper-oxide
  thin films synthesized by molecular beam epitaxy},\ }\href@noop {} {\bibfield
   {journal} {\bibinfo  {journal} {J. Electron Spectrosc. Relat. Phenom.}\
  }\textbf {\bibinfo {volume} {257}},\ \bibinfo {pages} {146775} (\bibinfo
  {year} {2022})}\BibitemShut {NoStop}%
\bibitem [{\citenamefont {Goler}\ \emph {et~al.}(2013)\citenamefont {Goler},
  \citenamefont {Coletti}, \citenamefont {Piazza}, \citenamefont {Pingue},
  \citenamefont {Colangelo}, \citenamefont {Pellegrini}, \citenamefont
  {Emtsev}, \citenamefont {Forti}, \citenamefont {Starke}, \citenamefont
  {Beltram},\ and\ \citenamefont {Heun}}]{GOLER2013249}%
  \BibitemOpen
  \bibfield  {author} {\bibinfo {author} {\bibfnamefont {S.}~\bibnamefont
  {Goler}}, \bibinfo {author} {\bibfnamefont {C.}~\bibnamefont {Coletti}},
  \bibinfo {author} {\bibfnamefont {V.}~\bibnamefont {Piazza}}, \bibinfo
  {author} {\bibfnamefont {P.}~\bibnamefont {Pingue}}, \bibinfo {author}
  {\bibfnamefont {F.}~\bibnamefont {Colangelo}}, \bibinfo {author}
  {\bibfnamefont {V.}~\bibnamefont {Pellegrini}}, \bibinfo {author}
  {\bibfnamefont {K.~V.}\ \bibnamefont {Emtsev}}, \bibinfo {author}
  {\bibfnamefont {S.}~\bibnamefont {Forti}}, \bibinfo {author} {\bibfnamefont
  {U.}~\bibnamefont {Starke}}, \bibinfo {author} {\bibfnamefont
  {F.}~\bibnamefont {Beltram}},\ and\ \bibinfo {author} {\bibfnamefont
  {S.}~\bibnamefont {Heun}},\ }\bibfield  {title} {\bibinfo {title} {Revealing
  the atomic structure of the buffer layer between sic(0001) and epitaxial
  graphene},\ }\href
  {https://doi.org/https://doi.org/10.1016/j.carbon.2012.08.050} {\bibfield
  {journal} {\bibinfo  {journal} {Carbon}\ }\textbf {\bibinfo {volume} {51}},\
  \bibinfo {pages} {249} (\bibinfo {year} {2013})}\BibitemShut {NoStop}%
\bibitem [{\citenamefont {Zhang}\ \emph {et~al.}(2008)\citenamefont {Zhang},
  \citenamefont {Brar}, \citenamefont {Wang}, \citenamefont {Girit},
  \citenamefont {Yayon}, \citenamefont {Panlasigui}, \citenamefont {Zettl},\
  and\ \citenamefont {Crommie}}]{zhang_giant_2008}%
  \BibitemOpen
  \bibfield  {author} {\bibinfo {author} {\bibfnamefont {Y.}~\bibnamefont
  {Zhang}}, \bibinfo {author} {\bibfnamefont {V.~W.}\ \bibnamefont {Brar}},
  \bibinfo {author} {\bibfnamefont {F.}~\bibnamefont {Wang}}, \bibinfo {author}
  {\bibfnamefont {C.}~\bibnamefont {Girit}}, \bibinfo {author} {\bibfnamefont
  {Y.}~\bibnamefont {Yayon}}, \bibinfo {author} {\bibfnamefont
  {M.}~\bibnamefont {Panlasigui}}, \bibinfo {author} {\bibfnamefont
  {A.}~\bibnamefont {Zettl}},\ and\ \bibinfo {author} {\bibfnamefont {M.~F.}\
  \bibnamefont {Crommie}},\ }\bibfield  {title} {\bibinfo {title} {Giant
  phonon-induced conductance in scanning tunnelling spectroscopy of
  gate-tunable graphene},\ }\href {https://doi.org/10.1038/nphys1022}
  {\bibfield  {journal} {\bibinfo  {journal} {Nature Physics}\ }\textbf
  {\bibinfo {volume} {4}},\ \bibinfo {pages} {627} (\bibinfo {year}
  {2008})}\BibitemShut {NoStop}%
\bibitem [{\citenamefont {Reich}\ \emph {et~al.}(2002)\citenamefont {Reich},
  \citenamefont {Maultzsch}, \citenamefont {Thomsen},\ and\ \citenamefont
  {Ordej\'on}}]{Reich2002}%
  \BibitemOpen
  \bibfield  {author} {\bibinfo {author} {\bibfnamefont {S.}~\bibnamefont
  {Reich}}, \bibinfo {author} {\bibfnamefont {J.}~\bibnamefont {Maultzsch}},
  \bibinfo {author} {\bibfnamefont {C.}~\bibnamefont {Thomsen}},\ and\ \bibinfo
  {author} {\bibfnamefont {P.}~\bibnamefont {Ordej\'on}},\ }\bibfield  {title}
  {\bibinfo {title} {Tight-binding description of graphene},\ }\href
  {https://doi.org/10.1103/PhysRevB.66.035412} {\bibfield  {journal} {\bibinfo
  {journal} {Phys. Rev. B}\ }\textbf {\bibinfo {volume} {66}},\ \bibinfo
  {pages} {035412} (\bibinfo {year} {2002})}\BibitemShut {NoStop}%
\bibitem [{\citenamefont {Ku}\ \emph {et~al.}(2010)\citenamefont {Ku},
  \citenamefont {Berlijn},\ and\ \citenamefont {Lee}}]{Wei2010}%
  \BibitemOpen
  \bibfield  {author} {\bibinfo {author} {\bibfnamefont {W.}~\bibnamefont
  {Ku}}, \bibinfo {author} {\bibfnamefont {T.}~\bibnamefont {Berlijn}},\ and\
  \bibinfo {author} {\bibfnamefont {C.-C.}\ \bibnamefont {Lee}},\ }\bibfield
  {title} {\bibinfo {title} {Unfolding first-principles band structures},\
  }\href {https://doi.org/10.1103/PhysRevLett.104.216401} {\bibfield  {journal}
  {\bibinfo  {journal} {Phys. Rev. Lett.}\ }\textbf {\bibinfo {volume} {104}},\
  \bibinfo {pages} {216401} (\bibinfo {year} {2010})}\BibitemShut {NoStop}%
\bibitem [{\citenamefont {Popescu}\ and\ \citenamefont
  {Zunger}(2012)}]{Popescu2012}%
  \BibitemOpen
  \bibfield  {author} {\bibinfo {author} {\bibfnamefont {V.}~\bibnamefont
  {Popescu}}\ and\ \bibinfo {author} {\bibfnamefont {A.}~\bibnamefont
  {Zunger}},\ }\bibfield  {title} {\bibinfo {title} {Extracting {$E$} versus
  $\vec{k}$ effective band structure from supercell calculations on alloys and
  impurities},\ }\href {https://doi.org/10.1103/PhysRevB.85.085201} {\bibfield
  {journal} {\bibinfo  {journal} {Phys. Rev. B}\ }\textbf {\bibinfo {volume}
  {85}},\ \bibinfo {pages} {085201} (\bibinfo {year} {2012})}\BibitemShut
  {NoStop}%
\bibitem [{\citenamefont {Farjam}(2015)}]{Farjam2015}%
  \BibitemOpen
  \bibfield  {author} {\bibinfo {author} {\bibfnamefont {M.}~\bibnamefont
  {Farjam}},\ }\href {https://doi.org/10.48550/ARXIV.1504.04937} {\bibinfo
  {title} {Projection operator approach to unfolding supercell band
  structures}} (\bibinfo {year} {2015})\BibitemShut {NoStop}%
\bibitem [{\citenamefont {Willke}\ \emph {et~al.}(2015)\citenamefont {Willke},
  \citenamefont {Amani}, \citenamefont {Sinterhauf}, \citenamefont {Thakur},
  \citenamefont {Kotzott}, \citenamefont {Druga}, \citenamefont {Weikert},
  \citenamefont {Maiti}, \citenamefont {Hofsäss},\ and\ \citenamefont
  {Wenderoth}}]{willke_doping_2015}%
  \BibitemOpen
  \bibfield  {author} {\bibinfo {author} {\bibfnamefont {P.}~\bibnamefont
  {Willke}}, \bibinfo {author} {\bibfnamefont {J.~A.}\ \bibnamefont {Amani}},
  \bibinfo {author} {\bibfnamefont {A.}~\bibnamefont {Sinterhauf}}, \bibinfo
  {author} {\bibfnamefont {S.}~\bibnamefont {Thakur}}, \bibinfo {author}
  {\bibfnamefont {T.}~\bibnamefont {Kotzott}}, \bibinfo {author} {\bibfnamefont
  {T.}~\bibnamefont {Druga}}, \bibinfo {author} {\bibfnamefont
  {S.}~\bibnamefont {Weikert}}, \bibinfo {author} {\bibfnamefont
  {K.}~\bibnamefont {Maiti}}, \bibinfo {author} {\bibfnamefont
  {H.}~\bibnamefont {Hofsäss}},\ and\ \bibinfo {author} {\bibfnamefont
  {M.}~\bibnamefont {Wenderoth}},\ }\bibfield  {title} {\bibinfo {title}
  {Doping of {Graphene} by {Low}-{Energy} {Ion} {Beam} {Implantation}:
  {Structural}, {Electronic}, and {Transport} {Properties}},\ }\href
  {https://doi.org/10.1021/acs.nanolett.5b01280} {\bibfield  {journal}
  {\bibinfo  {journal} {Nano Letters}\ }\textbf {\bibinfo {volume} {15}},\
  \bibinfo {pages} {5110} (\bibinfo {year} {2015})},\ \bibinfo {note}
  {publisher: American Chemical Society}\BibitemShut {NoStop}%
\bibitem [{\citenamefont {Bertóti}\ \emph {et~al.}(2015)\citenamefont
  {Bertóti}, \citenamefont {Mohai},\ and\ \citenamefont
  {László}}]{BERTOTI2015185}%
  \BibitemOpen
  \bibfield  {author} {\bibinfo {author} {\bibfnamefont {I.}~\bibnamefont
  {Bertóti}}, \bibinfo {author} {\bibfnamefont {M.}~\bibnamefont {Mohai}},\
  and\ \bibinfo {author} {\bibfnamefont {K.}~\bibnamefont {László}},\
  }\bibfield  {title} {\bibinfo {title} {Surface modification of graphene and
  graphite by nitrogen plasma: Determination of chemical state alterations and
  assignments by quantitative x-ray photoelectron spectroscopy},\ }\href
  {https://doi.org/https://doi.org/10.1016/j.carbon.2014.11.056} {\bibfield
  {journal} {\bibinfo  {journal} {Carbon}\ }\textbf {\bibinfo {volume} {84}},\
  \bibinfo {pages} {185} (\bibinfo {year} {2015})}\BibitemShut {NoStop}%
\bibitem [{\citenamefont {Rybin}\ \emph {et~al.}(2016)\citenamefont {Rybin},
  \citenamefont {Pereyaslavtsev}, \citenamefont {Vasilieva}, \citenamefont
  {Myasnikov}, \citenamefont {Sokolov}, \citenamefont {Pavlova}, \citenamefont
  {Obraztsova}, \citenamefont {Khomich}, \citenamefont {Ralchenko},\ and\
  \citenamefont {Obraztsova}}]{RYBIN2016196}%
  \BibitemOpen
  \bibfield  {author} {\bibinfo {author} {\bibfnamefont {M.}~\bibnamefont
  {Rybin}}, \bibinfo {author} {\bibfnamefont {A.}~\bibnamefont
  {Pereyaslavtsev}}, \bibinfo {author} {\bibfnamefont {T.}~\bibnamefont
  {Vasilieva}}, \bibinfo {author} {\bibfnamefont {V.}~\bibnamefont
  {Myasnikov}}, \bibinfo {author} {\bibfnamefont {I.}~\bibnamefont {Sokolov}},
  \bibinfo {author} {\bibfnamefont {A.}~\bibnamefont {Pavlova}}, \bibinfo
  {author} {\bibfnamefont {E.}~\bibnamefont {Obraztsova}}, \bibinfo {author}
  {\bibfnamefont {A.}~\bibnamefont {Khomich}}, \bibinfo {author} {\bibfnamefont
  {V.}~\bibnamefont {Ralchenko}},\ and\ \bibinfo {author} {\bibfnamefont
  {E.}~\bibnamefont {Obraztsova}},\ }\bibfield  {title} {\bibinfo {title}
  {Efficient nitrogen doping of graphene by plasma treatment},\ }\href
  {https://doi.org/https://doi.org/10.1016/j.carbon.2015.09.056} {\bibfield
  {journal} {\bibinfo  {journal} {Carbon}\ }\textbf {\bibinfo {volume} {96}},\
  \bibinfo {pages} {196} (\bibinfo {year} {2016})}\BibitemShut {NoStop}%
\bibitem [{\citenamefont {Wei}\ \emph {et~al.}(2009)\citenamefont {Wei},
  \citenamefont {Liu}, \citenamefont {Wang}, \citenamefont {Zhang},
  \citenamefont {Huang},\ and\ \citenamefont {Yu}}]{wei_synthesis_2009}%
  \BibitemOpen
  \bibfield  {author} {\bibinfo {author} {\bibfnamefont {D.}~\bibnamefont
  {Wei}}, \bibinfo {author} {\bibfnamefont {Y.}~\bibnamefont {Liu}}, \bibinfo
  {author} {\bibfnamefont {Y.}~\bibnamefont {Wang}}, \bibinfo {author}
  {\bibfnamefont {H.}~\bibnamefont {Zhang}}, \bibinfo {author} {\bibfnamefont
  {L.}~\bibnamefont {Huang}},\ and\ \bibinfo {author} {\bibfnamefont
  {G.}~\bibnamefont {Yu}},\ }\bibfield  {title} {\bibinfo {title} {Synthesis of
  {N}-{Doped} {Graphene} by {Chemical} {Vapor} {Deposition} and {Its}
  {Electrical} {Properties}},\ }\href {https://doi.org/10.1021/nl803279t}
  {\bibfield  {journal} {\bibinfo  {journal} {Nano Letters}\ }\textbf {\bibinfo
  {volume} {9}},\ \bibinfo {pages} {1752} (\bibinfo {year} {2009})},\ \bibinfo
  {note} {publisher: American Chemical Society}\BibitemShut {NoStop}%
\bibitem [{\citenamefont {Schiros}\ \emph {et~al.}(2012)\citenamefont
  {Schiros}, \citenamefont {Nordlund}, \citenamefont {Pálová}, \citenamefont
  {Prezzi}, \citenamefont {Zhao}, \citenamefont {Kim}, \citenamefont
  {Wurstbauer}, \citenamefont {Gutiérrez}, \citenamefont {Delongchamp},
  \citenamefont {Jaye}, \citenamefont {Fischer}, \citenamefont {Ogasawara},
  \citenamefont {Pettersson}, \citenamefont {Reichman}, \citenamefont {Kim},
  \citenamefont {Hybertsen},\ and\ \citenamefont
  {Pasupathy}}]{schiros_connecting_2012}%
  \BibitemOpen
  \bibfield  {author} {\bibinfo {author} {\bibfnamefont {T.}~\bibnamefont
  {Schiros}}, \bibinfo {author} {\bibfnamefont {D.}~\bibnamefont {Nordlund}},
  \bibinfo {author} {\bibfnamefont {L.}~\bibnamefont {Pálová}}, \bibinfo
  {author} {\bibfnamefont {D.}~\bibnamefont {Prezzi}}, \bibinfo {author}
  {\bibfnamefont {L.}~\bibnamefont {Zhao}}, \bibinfo {author} {\bibfnamefont
  {K.~S.}\ \bibnamefont {Kim}}, \bibinfo {author} {\bibfnamefont
  {U.}~\bibnamefont {Wurstbauer}}, \bibinfo {author} {\bibfnamefont
  {C.}~\bibnamefont {Gutiérrez}}, \bibinfo {author} {\bibfnamefont
  {D.}~\bibnamefont {Delongchamp}}, \bibinfo {author} {\bibfnamefont
  {C.}~\bibnamefont {Jaye}}, \bibinfo {author} {\bibfnamefont {D.}~\bibnamefont
  {Fischer}}, \bibinfo {author} {\bibfnamefont {H.}~\bibnamefont {Ogasawara}},
  \bibinfo {author} {\bibfnamefont {L.~G.~M.}\ \bibnamefont {Pettersson}},
  \bibinfo {author} {\bibfnamefont {D.~R.}\ \bibnamefont {Reichman}}, \bibinfo
  {author} {\bibfnamefont {P.}~\bibnamefont {Kim}}, \bibinfo {author}
  {\bibfnamefont {M.~S.}\ \bibnamefont {Hybertsen}},\ and\ \bibinfo {author}
  {\bibfnamefont {A.~N.}\ \bibnamefont {Pasupathy}},\ }\bibfield  {title}
  {\bibinfo {title} {Connecting {Dopant} {Bond} {Type} with {Electronic}
  {Structure} in {N}-{Doped} {Graphene}},\ }\href
  {https://doi.org/10.1021/nl301409h} {\bibfield  {journal} {\bibinfo
  {journal} {Nano Letters}\ }\textbf {\bibinfo {volume} {12}},\ \bibinfo
  {pages} {4025} (\bibinfo {year} {2012})},\ \bibinfo {note} {publisher:
  American Chemical Society}\BibitemShut {NoStop}%
\bibitem [{\citenamefont {Wang}\ \emph {et~al.}(2013)\citenamefont {Wang},
  \citenamefont {Wei}, \citenamefont {Jin}, \citenamefont {Ning}, \citenamefont
  {Yu}, \citenamefont {Fu},\ and\ \citenamefont
  {Bao}}]{wang_simultaneous_2013}%
  \BibitemOpen
  \bibfield  {author} {\bibinfo {author} {\bibfnamefont {Z.-j.}\ \bibnamefont
  {Wang}}, \bibinfo {author} {\bibfnamefont {M.}~\bibnamefont {Wei}}, \bibinfo
  {author} {\bibfnamefont {L.}~\bibnamefont {Jin}}, \bibinfo {author}
  {\bibfnamefont {Y.}~\bibnamefont {Ning}}, \bibinfo {author} {\bibfnamefont
  {L.}~\bibnamefont {Yu}}, \bibinfo {author} {\bibfnamefont {Q.}~\bibnamefont
  {Fu}},\ and\ \bibinfo {author} {\bibfnamefont {X.}~\bibnamefont {Bao}},\
  }\bibfield  {title} {\bibinfo {title} {Simultaneous {N}-intercalation and
  {N}-doping of epitaxial graphene on {6H}-{SiC}(0001) through thermal
  reactions with ammonia},\ }\href {https://doi.org/10.1007/s12274-013-0317-7}
  {\bibfield  {journal} {\bibinfo  {journal} {Nano Research}\ }\textbf
  {\bibinfo {volume} {6}},\ \bibinfo {pages} {399} (\bibinfo {year}
  {2013})}\BibitemShut {NoStop}%
\bibitem [{\citenamefont {Du}\ \emph {et~al.}(2023)\citenamefont {Du},
  \citenamefont {Li}, \citenamefont {Gu}, \citenamefont {Pasupathy},
  \citenamefont {Tranquada},\ and\ \citenamefont
  {Fujita}}]{PhysRevX.13.021025}%
  \BibitemOpen
  \bibfield  {author} {\bibinfo {author} {\bibfnamefont {Z.}~\bibnamefont
  {Du}}, \bibinfo {author} {\bibfnamefont {H.}~\bibnamefont {Li}}, \bibinfo
  {author} {\bibfnamefont {G.}~\bibnamefont {Gu}}, \bibinfo {author}
  {\bibfnamefont {A.~N.}\ \bibnamefont {Pasupathy}}, \bibinfo {author}
  {\bibfnamefont {J.~M.}\ \bibnamefont {Tranquada}},\ and\ \bibinfo {author}
  {\bibfnamefont {K.}~\bibnamefont {Fujita}},\ }\bibfield  {title} {\bibinfo
  {title} {Periodic atomic displacements and visualization of the
  electron-lattice interaction in the cuprate},\ }\href
  {https://doi.org/10.1103/PhysRevX.13.021025} {\bibfield  {journal} {\bibinfo
  {journal} {Phys. Rev. X}\ }\textbf {\bibinfo {volume} {13}},\ \bibinfo
  {pages} {021025} (\bibinfo {year} {2023})}\BibitemShut {NoStop}%
\bibitem [{\citenamefont {Godfrin}(1991)}]{Godfrin_1991}%
  \BibitemOpen
  \bibfield  {author} {\bibinfo {author} {\bibfnamefont {E.~M.}\ \bibnamefont
  {Godfrin}},\ }\bibfield  {title} {\bibinfo {title} {A method to compute the
  inverse of an n-block tridiagonal quasi-hermitian matrix},\ }\href
  {https://doi.org/10.1088/0953-8984/3/40/005} {\bibfield  {journal} {\bibinfo
  {journal} {Journal of Physics: Condensed Matter}\ }\textbf {\bibinfo {volume}
  {3}},\ \bibinfo {pages} {7843} (\bibinfo {year} {1991})}\BibitemShut
  {NoStop}%
\end{thebibliography}%


\providecommand{\noopsort}[1]{}\providecommand{\singleletter}[1]{#1}
%
\end{document}